%% file: Camera_ToN.tex
\documentclass[10pt, journal]{IEEEtran}
\usepackage{xspace}
\usepackage{threeparttable}
\usepackage{color}
\usepackage{url}
\usepackage{subcaption}
\usepackage{paralist}
\usepackage{wrapfig}
\usepackage{multirow}
\usepackage{listings}
\usepackage{booktabs}
\usepackage{makecell}
\usepackage{boxedminipage}
\usepackage{amsmath}
\usepackage{graphicx}
\usepackage{minibox}
\usepackage{algorithm}
\usepackage{pifont}
\usepackage{etoolbox}
\usepackage{comment}

\usepackage{amssymb}
\usepackage{color}
\usepackage{xcolor}
\usepackage{subfiles}
\usepackage[utf8]{inputenc}
\usepackage[english]{babel}
\usepackage[cachedir=.]{minted}
\usepackage{xcolor}
\usepackage{ulem}
\usepackage{adjustbox}
\usepackage{diagbox}
\usepackage{colortbl}
\usepackage{balance} 
\usepackage{multicol}
\usepackage{nicefrac}

\newcommand{\ra}[1]{\renewcommand{\arraystretch}{#1}}

\AtBeginDocument{%
  \providecommand\BibTeX{{%
    \normalfont B\kern-0.5em{\scshape i\kern-0.25em b}\kern-0.8em\TeX}}}

\input{macro.tex}

\begin{document}

\title{Leader Rotation Is Not Enough: Scrutinizing Leadership Democracy of Chained BFT Consensus}

\author{
     Jianyu Niu,~\IEEEmembership{Member, ~IEEE,} Yining Tang, Runchao Han, Chen Feng,~\IEEEmembership{Member, ~IEEE,} \\
    Yinqian Zhang,~\IEEEmembership{Member, ~IEEE} 
    
    \IEEEcompsocitemizethanks{
        
        Jianyu Niu, Yining Tang, and Yinqian Zhang are with the Research Institute of Trustworthy Autonomous Systems and the Department of Computer Science and Engineering, Southern University of Science and Technology, Shenzhen, China.
        Email: tangyn2018@mail.sustech.edu.cn, niujy@sustech.edu.cn and yinqianz@acm.org. 
        
        Runchao Han is with Babylon Labs. E-mail: runchao@babylonlabs.io.
        
        Chen Feng is with Blockchain@UBC and the School of Engineering, The University of British Columbia (Okanagan Campus), Kelowna, BC, Canada. Email: chen.feng@ubc.ca. 

    }
}

\maketitle

\begin{abstract} 
With the growing popularity of blockchains, modern chained BFT protocols combining chaining and leader rotation to obtain better efficiency and leadership democracy have received increasing interest. Although the efficiency provisions of chained BFT protocols have been thoroughly analyzed, the leadership democracy has received little attention in prior work. 
In this paper, we scrutinize the leadership democracy of four representative chained BFT protocols, especially under attack. 
To this end, we propose a unified framework with two evaluation metrics, \ie, chain quality and censorship resilience, and quantitatively analyze chosen protocols through the Markov Decision Process (MDP). 
With this framework, we further examine the impact of two key components, \ie, voting pattern and leader rotation, on leadership democracy.   
Our results indicate that leader rotation is not enough to provide the leadership democracy guarantee; an adversary could utilize the design, \eg, voting pattern, to deteriorate the leadership democracy significantly. 
Based on the analysis results, we propose customized countermeasures for three evaluated protocols to improve their leadership democracy with only slight protocol overhead and no change of consensus rules.
We also discuss future directions toward building more democratic chained BFT protocols. 
\end{abstract}


\section{Introduction} \label{sec:intro}
The rising prominence of decentralized applications such as global payment~\cite{nakamoto2012bitcoin, zcash2014} and Decentralized Finance (DeFi)~\cite{ethereum} has renewed the interest in Byzantine Fault Tolerant (BFT) consensus---the backbone of blockchains and further spawned a plethora of new protocols~\cite{gilad2017algorand, GLT20, MXC16, Buchman2016TendermintBF, casper, hotstuffPODC, team2021diembft, shi2019streamlined, fastHotStuff, malkhi2023HotStuff, giridharan2021no, sui2022marlin, gelashvili2022jolteon, giridharan2023beegees}. Among them, a family of chained BFT protocols combining \textit{chaining} and \textit{leader rotation} has received increasing attention~\cite{casper, hotstuffPODC, team2021diembft, shi2019streamlined, fastHotStuff, malkhi2023HotStuff, giridharan2021no, sui2022marlin, gelashvili2022jolteon,
giridharan2023beegees}.
Prominent examples are Casper FFG~\cite{casper}, chained HotStuff~\cite{hotstuffPODC}, Streamlet~\cite{shi2019streamlined}, Fast-HotStuff~\cite{fastHotStuff}, BeeGees~\cite{giridharan2023beegees}, etc. 
Moreover, these protocols have been used in tens of blockchains, including both 
permissioned ones (\eg, XuperChain~\cite{baidu} and 
Hyperchain~\cite{Hyperchain}) and permissionless ones (\eg, Ethereum 2.0~\cite{schwarz2022three}, Aptos~\cite{aptos}, Cypherium~\cite{cypherium}, Flow~\cite{hentschel2002flow}, Zilliqa 2.0~\cite{zilliqa}, and DeSo~\cite{Deso}). 


One reason for the popularity of chained BFT protocols in blockchains, especially permissionless ones, is that they can obtain better \textit{efficiency} and \textit{leadership democracy} (also known as fairness~\cite{abrahamOPODIS}) than classic BFT protocols~\cite{abrahamOPODIS, giridharan2023beegees}.
First, chained BFT consensus runs in views, with each having a delegated node, \ie, leader, to propose a block linked with a previous block and then collect enough nodes' votes to establish an associated quorum certificate (QC). 
The \textit{chaining block structure} can pipeline the voting phases of consecutive blocks to avoid redundant message transmission and verification, resulting in high efficiency.
Second, in chained BFT, the leader is rotated for each view and so has at most one proposed block to be committed in one view. 
This frequent rotation is also known as democracy-favoring leadership policy~\cite{shi2019streamlined, chanpili, Shrestha20}. 
In contrast, classic BFT protocols (\eg, PBFT~\cite{castro1999practical}) adopt a stability-favoring policy where a leader is only replaced if it fails to make consensus progress in a period. 
Therefore, chained BFT protocols allow every node to \textit{fairly} propose blocks, resulting in better leadership democracy. 

However, many recent studies~\cite{giridharan2023beegees, niu2021performance, gai2021dissecting, dinh2017blockbench, shapiro2020performance, amiri2022bedrock, gramoli2023diablo} have shown that Byzantine nodes (\ie, behaving arbitrarily) can significantly deteriorate the system efficiency in terms of longer latency and lower throughput. 
For example, Cohen~\etal~\cite{cohen2022aware} shows that the throughput of the chained HotStuff (or CHS in short) can drop over 30x and the latency increases 5x under attack in a setting of 1-3 Byzantine nodes out of 10 nodes.  
Even worse, Giridharan~\etal~\cite{giridharan2023beegees} reveal an attack case to violate the liveness of CHS, \ie, making no block committed.  
The root cause for these attacks is chaining, a block is not committed directly, but rather committed after receiving some blocks with QCs produced in consecutive views (\ssecref{sec:background}). Thus, when a Byzantine node is elected as the leader, it can strategically deviate from the protocol (\eg, proposing no block~\cite{cohen2022aware} or never broadcasting the QC of the previous block) to break the commitment condition of blocks. 

Although efficiency provisions of chained BFT protocols have been well-studied, leadership democracy has received little attention in prior work. 
Leadership democracy is important for blockchains, since a leader decides transaction ordering, allowing them to gain an advantage in decentralized applications like auctions and cryptocurrency exchanges, or obtain protocol rewards. 
As such, the concept of \textit{leadership democracy}, which we coin to denote the fair opportunity for any node to become a leader and propose outstanding transactions in blocks, assumes paramount significance~\cite{eyal2018majority, niu2019selfish, Pass2017fruit, fairness}. 
This is also stated as the basic premise of chained BFT protocols~\cite{abrahamOPODIS, giridharan2023beegees, cohen2022aware, Simplex}. 
However, the revealed degraded efficiency of chained BFT protocols under attack casts a hint of worry about the leadership democracy premise, making us wonder: \textit{To what extent can chained BFT protocols guarantee leadership democracy under attack?} 

In this paper, we aim to answer this question by performing the first comprehensive evaluation of leadership democracy in chained BFT protocols. 
Our work advocates more attention to leadership democracy and the importance of systematical evaluation to explore potential attacks, understand the impact of design components, and make fair comparisons of chained BFT protocols. 
Our contributions include:

\bheading{Building a generic evaluation framework (\ssecref{sec:model} and \ssecref{sec:MDPmodel}).} We first introduce an evaluation framework with two metrics: \textit{chain quality} and \textit{censorship resilience} (\ssecref{subsec:metrics}). Chain quality is defined as the fraction of committed blocks proposed by honest nodes, while censorship resilience captures the difficulty for an adversary to censor blocks proposed by certain honest nodes. They not only represent the premise of fairness and censorship resilience for chained BFT protocol~\cite{abrahamOPODIS}, but also reflect the security issues associated with leadership democracy concerned by the community~\cite{garay2015bitcoin, kiayias2017ouroboros, Zhang2019CommonMetrics, huang2021rich} (see more in \ssecref{subsec:issues}).

Within the framework, we consider possible adversarial behaviors on chained BFT protocols and explore the optimal attack strategy to minimize the two metrics. This exploration differs from the existing analysis that only considers some deterministic strategies~\cite{giridharan2023beegees, niu2021performance, gai2021dissecting, dinh2017blockbench, shapiro2020performance, amiri2022bedrock, gramoli2023diablo}. 
It also enables us to compare different chained BFT protocols fairly. 
To this end, we utilize Markov Decision Process (MDP) to explore the optimal adversarial strategy (\ssecref{subsec:mdpdesign}). 
MDPs can accurately capture the sequential and probabilistic nature of the block proposing process, allowing us to find utility-maximizing strategies in stochastic environments. Unlike game theory or discrete event simulations, MDPs efficiently handle dynamic state transitions and offer flexibility to adapt the model to different blockchain protocols and attack scenarios. 

However, chained BFT protocols differ in numerous design components, making it difficult to model them in MDP. 
Thus, we identify several key components (\ie, consensus rules, voting patterns, and leader rotation) that affect leadership democracy.  
Besides, we properly simplify the chosen components and confine the adversary’s actions to build a general MDP model for various protocols. 
Although prior works have used MDP to analyze Nakamoto-style consensus~\cite{sapirshtein2017optimal, gervais2016security, zhang2017necessity}, they cannot be applied to chained BFT consensus because of the notable differences in consensus rules. 
For example, BFT has (explicit) voting and provides finality~\cite{gilad2017algorand}, whereas NC-style has no (implicit) voting pattern and only provides probabilistic security.

\bheading{Quantifying and comparing the optimality gap among chained BFT protocols (\ssecref{sec:Protocolanalysis}).}  We evaluate the leadership democracy of four pioneering chained BFT protocols: Two-chain HotStuff (denoted as 2CHS), CHS, Streamlet, and Fast-HotStuff (denoted as FHS). 
They have been widely adopted in the industry~\cite{aptos, cypherium, baidu, Hotshot} and also inspired many subsequent chained BFT protocols~\cite{danezis2022narwhal, Kauri, gai2023scaling, sheng2021bft, stathakopoulou2022state, neu2021ebb, decouchant2022damysus}. (See more choice reasons in \ssecref{sec:related}.)
Our evaluation results show that none can achieve the optimal values in the two metrics. Specifically, CHS has the worst chain quality and censorship resilience under the same fraction of Byzantine nodes. 
For example, the chain quality (resp., censorship resilience) of CHS decreases by $0.29$ (resp., $0.56$) from the optimal value when the fraction of Byzantine nodes is $1/3$. 

By observing the optimal strategies for these protocols, we find that an adversary can always prevent some blocks proposed by honest nodes from being committed without any risk of its blocks. In other words, the attacks on these protocols are riskless. 
It also implies that even a single Byzantine node can obtain unfair chain quality from these protocols. 
The root reason for this is the deterministic nature of rules; when a node tries to fork a proposed block, it can always make its block win. 
Besides, the strategies also identify some new attacks in 
Streamlet and FHS. 
For example, the adversary can strategically release its withheld block to override subsequent blocks from honest nodes in Streamlet, while hiding the QC of previous blocks to create forks in FHS. Besides, we also validate the effect of the obtained optimal strategies on chosen protocols in the open-source benchmark platform Bamboo~\cite{gai2021dissecting}.

\bheading{Exploring lightweight countermeasures (\ssecref{sec:countermeasure}).} According to the found vulnerabilities and weaknesses in the analysis, we propose specific countermeasures for three protocols. 
In particular, our countermeasures should not \textit{modify voting and committing rules}, which ensures that the safety of chained BFT protocols is not affected. 
Guided by this principle, our countermeasures include optimizing the \textit{voting pattern} in FHS and introducing randomness in the \textit{proposing rule} of 2CHS and CHS.
Our evaluation results illustrate that these countermeasures can significantly improve leadership democracy under attack. We also identified the difficulty of thwarting attacks in Streamlet without breaking the principle. 
Some key points are summarized as follows:

Our countermeasures suggest that even minor changes in design components (\eg, \textit{voting pattern} in \ssecref{subsec:BackchainedBFT}) substantially improve leadership democracy. For example,  with our countermeasures, an adversary has to control more than $0.285$ of Byzantine nodes (instead of a single Byzantine node in CHS) to obtain an unfair chain quality in CHS. Besides, FHS with our countermeasure can achieve optimal chain quality and censorship resilience. 
Our countermeasures confirm security-performance trade-offs in designing chained BFT protocols. No enhanced version of these four protocols can simultaneously achieve linear message complexity, responsiveness 
, and optimal leadership democracy. 
Navigating the trade-offs becomes a pivotal aspect of decision-making in protocol design. Furthermore, we discuss future directions for chained BFT protocols with better leadership democracy in Appendix~\ref{appen:discussion}.

\section{Preliminaries of Chained BFT Consensus} \label{sec:background}
In this section, we briefly recap chained BFT protocols and the associated leadership democracy issues in the context of blockchains. We choose four representative and influential chained BFT protocols: Two-chain HotStuff (2CHS), chained HotStuff (CHS), Streamlet, and Fast-HotStuff (FHS). (See detailed reasons for choosing them in \ssecref{sec:related}.)
Due to space constraints, we refer readers to \appref{appen:chainedBFT} for detailed descriptions of these chained BFT protocols.

\subsection{Chained BFT Protocols} \label{subsec:BackchainedBFT}
\subsubsection{Overview of chained BFT} 
\begin{figure}[t]
    \centering
    \includegraphics[width=0.9\linewidth]{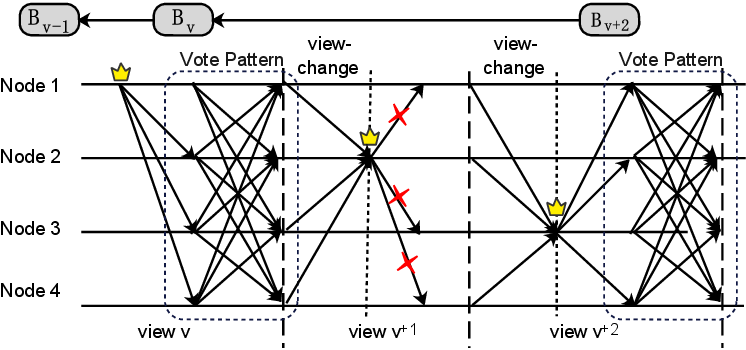}
    \caption{\textbf{The propose-vote paradigm of chained BFT protocols.} The yellow crown denotes the leader in a view.}
    \label{fig:paradigem}
\end{figure}

Chained BFT protocols allow a group of nodes to eventually reach an agreement on a chain of blocks and thus achieve a total ordering of transactions~\cite{giridharan2023beegees, gai2021dissecting}. Specifically, chained BFT protocols run in views, each with a single node, namely the leader, selected by certain \textit{election and rotation policies}, as shown in \figref{fig:paradigem}. The chaining structure enables chained BFT protocols to share a unified propose-vote paradigm to commit blocks by pipelining different consensus phases. This is, at the beginning of a view, the leader can propose a block containing a batch of transactions and a cryptographic reference of a previous block (also called parent block) chosen by the \texttt{proposing rule}. The included cryptographic references (\eg, hash value~\cite{shi2019streamlined, chan2020streamlet} or quorum certificate~\cite{hotstuffPODC}) link blocks into a chain structure. 
When receiving the block, nodes decide whether to update their local states and to vote the block according to the \texttt{voting rule}. These votes are collected in certain \textit{voting patterns} to form blocks' quorum certificates (QCs). Specifically, a block QC contains no less than $2f+1$ votes, and a block is certified by a node if there is a QC in the node's view. 

In a chained BFT protocol, a node uses the \texttt{committing rule} to decide whether a block is committed, given chaining certified blocks at its local state. For example, in Streamlet~\cite{chan2020streamlet}, a node commits the first two of three certified blocks produced in consecutive views. In particular, once a block is
committed, the entire prefix of the chain is also committed. 
Chained BFT protocols ensure that once a block is committed, no other block at the same height is also committed, \ie, the \textit{safety property}. 
To ensure the \textit{liveness property}, \ie, honest clients' transactions are eventually included in committed blocks, a chained BFT protocol has to advance its views if no certified block is produced in a certain period (due to the faulty leader or network partition). The mechanism to realize the view advancement is usually known as view-change in~\cite{fastHotStuff} or Pacemaker in~\cite{hotstuffPODC}. 

\subsubsection{Varied design components} \label{subsub:components}
Despite the same propose-vote paradigm, chained BFT protocols vary in multiple design components, \ie, the consensus rule, voting pattern, and leader election and rotation policy, to achieve different features such as linear message complexity, responsiveness~\cite{Thunderella}, etc. However, the impact of these design components on leadership democracy is unclear, which motivates us to scrutinize them. 

\bheading{Consensus rules.} The consensus rules consist of \texttt{proposing}, \texttt{voting}, and \texttt{committing} rules. Specifically, the \texttt{proposing rule} determines which parent block should be extended in a view; the \texttt{voting rule} decides whether a node can vote for the first-received block of the leader; the \texttt{committing rule} determines whether a block is committed. Generally speaking, the \texttt{voting rule} and \texttt{committing rule} jointly ensure the safety of chained BFT protocols, while the \texttt{proposing rule} guarantees that nodes will vote for the proposed block from the leader, \ie, the liveness property. 

Chained BFT protocols differ in consensus rules to achieve different characteristics. For example, compared with 2CHS, CHS adds one additional phase to commit blocks by revising the \texttt{committing rule} and \texttt{voting rule} to have responsiveness~\cite{hotstuffPODC}. The responsiveness property enables the system to reach consensus at the rate of the actual message delays rather than a known bounded message delay $\Delta$~\cite{Delaybounds}.
However, the additional phase increases the latency for committing a block. 
To reduce the overhead, FHS adds the highest QC proof in blocks and modifies the \texttt{voting rule} to achieve responsiveness.
Unlike them, Streamlet aims to make the protocol simple. Thus, its consensus rules are different from the other three protocols. More details are given in \appref{appen:chainedBFT}.

\begin{figure}[t]
    \centering
    \includegraphics[width=\linewidth]{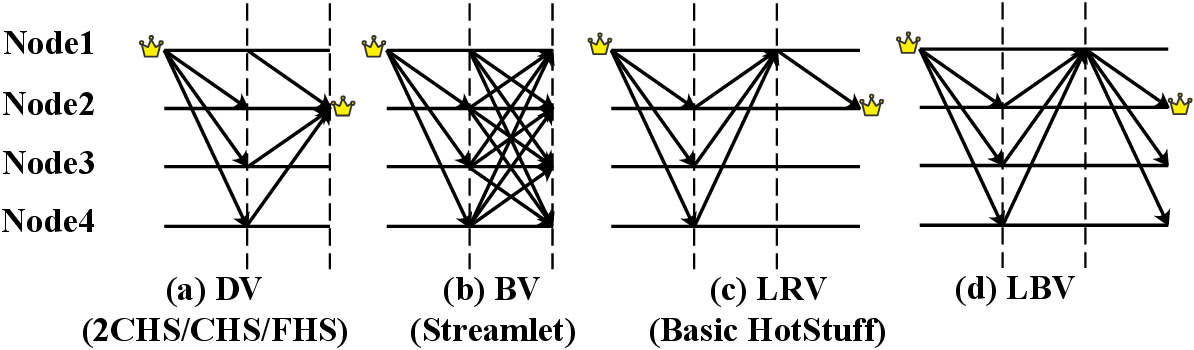}
    \caption{Four different voting patterns. Here, DV, BV, LRV, and LBV are short for ``direct votes'', ``broadcasting votes'', ``leader relay votes'' and ``leader broadcasting votes'', respectively. The yellow crown denotes the leader in a view.}
    \label{fig:msgptn}
\end{figure}

\bheading{Voting pattern.} 
Chained BFT protocols adopt different voting patterns to form and transmit a block's QC in a view. 
\figref{fig:msgptn} shows four voting patterns: $i$) direct votes (DV)~\cite{hotstuffPODC, team2021diembft, fastHotStuff}, in which nodes send votes directly to the next leader; $ii$) leader relay votes (LRV)~\cite{hotstuffArxiv}, in which nodes send votes to the current leader who then relays the QC to the next leader; $iii$) broadcasting votes (BV)~\cite{chan2020streamlet}, in which nodes broadcast their votes to each other; and $iv$) leader broadcast votes (LBV), in which nodes send votes to the current leader, who then broadcasts the QC to all. 
Different voting patterns have different message complexities (\ie, $O(n^2)$ in BV and $O(n)$ in others) and communication steps (\ie, two steps in DV and BV, while three steps in others). 

\bheading{Leader election and rotation.} 
Chained BFT protocols adopt different policies to select leaders. Two representative ones are the round-robin policy (\ie, nodes take turns to be the leader) and the random policy (\ie, a node is chosen to be the leader uniformly at random each time). 
When deployed in decentralized systems~\cite{schwarz2022three, aptos, cypherium, Hotshot}, a leader is usually randomly chosen from nodes according to their possession of some scarce resources (\eg, stake~\cite{schwarz2022three} or reputation~\cite{cohen2022aware}). 
In this paper, we focus on a random policy, where a leader is randomly chosen at the beginning of each view. We also briefly discuss the impact of the round-robin policy in Appendix~\ref{appen:robin}. 

\subsection{Forking Attack of Chained BFT} \label{subsub:forking}
Due to the chaining structure, a block in chained BFT protocols is not committed by the end of its generation view. As a result, it may not be extended by the block in the next view due to network delays or Byzantine leader behaviors~\cite{gai2021dissecting, niu2021performance, fastHotStuff}. 
In this case, these two conflicting blocks (\ie, no extending relationship) form a fork, and nodes observe a tree of blocks due to the forking.
To resolve forks, subsequent leaders follow the \texttt{proposing rule} to extend one of them. 
For example, a leader in CHS extends the parent block certified with the latest QC. Here, the freshness of QC is ranked by their generation views, and the latest QC has a higher view. 
With new blocks appended, one of their descendant blocks will be committed, making the associated block committed, too. 
In contrast, the uncommitted block will eventually be abandoned by all nodes. 
Specifically, the safety property ensures that honest nodes accept the same committed chain of blocks, referred to as the \textit{main chain}---a key concept that will be used in our analysis.
In other words, once a block is committed, it is included in the main chain. 

Previous studies~\cite{gai2021dissecting, niu2021performance, giridharan2023beegees, liu2023flexible} illustrate that an adversary in CHS can strategically create forks to significantly decrease the system throughput or increase latency. However, these works either focus on one certain protocol, lack comprehensive studies on chained BFT protocols, or do not examine them from the leadership democracy perspective. 
In this paper, we will systematically analyze chained BFT protocols under various attack strategies and examine their maximum impact on leadership democracy. 
Specifically, we are among the first to reveal new forking attacks on Fast-HotStuff and Streamlet. 

\subsection{Leadership Democracy} \label{subsec:issues}
Unlike Nakamoto-style consensus~\cite{nakamoto2012bitcoin} without requiring participants' identities, BFT consensus works in a setting of known participants and is usually used in permissioned blockchains. However, modern blockchains for realizing scalable, energy-efficient permissionless blockchains (\eg, Ethereum 2.0~\cite{casper} and Algorand~\cite{gilad2017algorand}) have already combined BFT consensus with Proof-of-Stake (PoS) (or Proof-of-Work (PoW)). Among existing adopted ones, chained BFT protocols utilizing chaining and leader rotation to obtain efficiency and leadership democracy have extensive interests. 
By now, multiple permissionless blockchains, \eg, Ethereum 2.0~\cite{schwarz2022three}, Aptos~\cite{aptos}, Cypherium~\cite{cypherium}, Flow~\cite{hentschel2002flow}, Zilliqa 2.0~\cite{zilliqa}, and DeSo~\cite{Deso}, have adopted them. More importantly, the increasing popularity shows the trend of their adoption on more upcoming platforms. 

Leadership democracy is one important premise for blockchains, especially permissionless ones (see \ssecref{sec:intro}). 
For example, Abraham~\etal~\cite{abrahamOPODIS} claim that ``leader rotation of chained BFT protocols is believed to bring better \textit{fairness}, \textit{censorship resistance}, ...". Unfortunately, the premise has not been checked in the prior work. 
To fill the gap, we focus on the mentioned two aspects: leadership fairness (\ie, fairness in proposing blocks) and block censorship, as introduced below.

\bheading{Leadership fairness.} 
Leadership fairness underscores the importance of providing nodes with equal opportunities to propose blocks and then order transactions. Thus, a \textit{fair} protocol should ensure that the proportion of committed blocks produced by honest nodes is proportional to the ratio of honest nodes~\cite{Pass2017fruit,Zhang2019CommonMetrics}.  
The fairness is important for permissionless blockchains, since a leader decides transaction ordering, allowing them to gain an advantage in decentralized applications like auctions and cryptocurrency exchanges. 
The importance also holds for permissioned blockchains; nodes may be inclined to solely serve their clients, and so leadership fairness facilitates the timely inclusion of clients' transactions in blocks. Furthermore, leadership fairness is closely related to reward fairness. To incentivize nodes' participation, blockchains usually provide certain rewards (\eg, block reward and transaction fees~\cite{nakamoto2012bitcoin, ethereum}) for leaders whose blocks are committed. 
Thus, unfair leadership may lead to various incentive attacks (\eg, selfish mining~\cite{eyal2018majority}). To capture leadership fairness, we introduce the widely used metric---chain quality~\cite{garay2015bitcoin, kiayias2017ouroboros, Zhang2019CommonMetrics, huang2021rich}, which is defined in~\ssecref{subsec:metrics}.  

\bheading{Block censorship.} 
Block censorship represents a significant threat within blockchain systems, where an adversary attempts to prevent some honest blocks from being committed. The forks created by the adversary often result in chain reorganization~\cite{d2022no, Zhang2019CommonMetrics, Tanniru2021}, leading to the abandonment of legitimate blocks proposed by honest leaders and the replacement of the main chain. This further excludes some certified transactions, thereby affecting the finality and ordering of the transactions.
By censoring certain blocks, the adversary can further launch attacks, \eg, front-running~\cite{eskandari2020sok}. 
Besides, the adversary can also strategically include or exclude transactions from their blocks to exploit the inherent value within the transaction data. Such attacks can lead to huge financial profits, which is usually known as Miner Extractable Value (MEV)~\cite{daian2020flash}. Given the profound impact of censorship, we introduce censorship resilience~\cite{garay2015bitcoin, kiayias2017ouroboros, Zhang2019CommonMetrics, huang2021rich}, which is defined in~\ssecref{subsec:metrics}.  

\section{System Model and Metrics} \label{sec:model}
\subsection{System Model}
We follow the system model of existing chained BFT protocols~\cite{casper, hotstuffPODC, team2021diembft, shi2019streamlined, fastHotStuff, malkhi2023HotStuff}.
There are $n= 3f+1$ nodes, in which at most $f$ nodes are \textit{Byzantine} (to ensure the safety and liveness of the system~\cite{castro1999practical, hotstuffPODC}). 
The \textit{Byzantine} nodes can behave arbitrarily, whereas the rest nodes are honest and strictly follow the protocol.
We assume the worst-case scenario where
all Byzantine nodes are controlled by a single adversary, referred to as ``\textit{the adversary}".  In other words, the adversary aims to maximize the effect of attacks by colluding all Byzantine nodes. 
In particular, to observe the impact of different numbers of Byzantine nodes, we use $t$ ($t \leq f$) to denote the number of Byzantine nodes among $n$ nodes in the system. 
We use $\alpha$ (resp., $\beta$) to denote the fraction of Byzantine (resp., honest) nodes among all nodes. We have $\alpha = t/n$ and $\beta = 1-\alpha$. 
We assume there is a leader election mechanism such that the probability of a Byzantine node (resp., honest node) being chosen as the leader is $\alpha$ (resp., $\beta$). We refer a leader from honest nodes (resp., Byzantine nodes) to an honest (resp., adversarial) leader.
Besides, we refer to a block produced by an honest leader (resp., the adversarial leader) as an honest (resp., adversarial) block. 

As for network connectivity, we assume nodes communicate through point-to-point, authenticated, and
reliable channels by following prior works~\cite{casper, hotstuffPODC, team2021diembft, shi2019streamlined, fastHotStuff, malkhi2023HotStuff}; the adversary cannot forge honest nodes' messages (\eg, blocks or votes) because it cannot forge the digital signatures of honest nodes. 
Besides, we follow the partially synchronous network model of chained BFT protocols; there is an unknown Global Stabilization Time (GST) and a known bound $\Delta$ such that one honest node's message will arrive at all other honest nodes within $\Delta$ after GST~\cite{casper, hotstuffPODC, team2021diembft, shi2019streamlined, fastHotStuff, malkhi2023HotStuff}. 
This paper focuses on analyzing chained BFT protocols during network synchrony (\ie, after GST) as prior analysis or evaluation works~\cite{niu2021performance, amiri2022bedrock, gramoli2023diablo, gai2021dissecting}.
We argue that during network asynchrony (\ie, before GST), the adversary can cause more damage to the leadership democracy by utilizing network partition. However, our analysis during network synchrony already reveals many insights about the design. We briefly discuss the impact of network asynchrony in Appendix~\ref{appen:asynchrnoy}. 
Since the studied metrics (introduced below) are time-agnostic, we do not consider the detailed message delay distributions. 

\begin{table}[t]
    \footnotesize
    \centering
    \setlength{\abovecaptionskip}{0cm}
    \caption{\textbf{Summary of Notations.}}
    \begin{tabular}{@{}m{0.25cm}l|m{0.25cm}l@{}}
        \toprule[1pt]
        Term    & Description  &  Term    & Description   \\
        \midrule
        $\alpha$  &  Byzantine node fraction &  $L$ & Flag of honest leader\\ 
        $\beta$  &  Honest node fraction & $l_a$  &  \# uncommited adversarial blocks \\ 
        $v$ & View number & $l_h$ & \# uncommited honest blocks \\ 
        $m$ & Number of views & ${B_a}$ & \# committed adversarial block  \\
        $Q(\alpha)$  & Chain quality & ${B_h}$ & \# committed honest block\\ 
        $C(\alpha)$ & Censorship resilience & ${O_h}$ & \# excluded honest block  \\
        \bottomrule[1pt]
    \end{tabular}
    \label{table:Notation}
    \vspace{-0.4cm}
\end{table}

\subsection{Evaluation Metrics} \label{subsec:metrics}
Leadership democracy of chained BFT protocols has not been examined in prior work, so we introduce a framework with two evaluation metrics: chain quality and censorship resilience. 
Specifically, chain quality has been widely used to measure the fairness for honest nodes to have block committed in permissionless blockchains~\cite{garay2015bitcoin, kiayias2017ouroboros, Zhang2019CommonMetrics, huang2021rich}, whereas censorship resilience denotes the difficulty for an adversary to censor certain honest nodes' blocks~\cite{Zhang2019CommonMetrics}. 
They further reflect the two leadership democracy issues concerning the blockchain community (\ssecref{subsec:issues}). Except for them, the proposed framework also offers the flexibility to integrate additional metrics as they arise within the community. {The notations used in this section are listed in \tabref{table:Notation}.}
 
\subsubsection{Chain quality} 
The chain quality $Q$ represents the fraction of honest blocks out of all blocks committed by the system, which further measures the difficulty for the adversary to substitute honest blocks from the main chain. 
In other words, by lowering the chain quality, the adversary can increase the fraction of adversarial blocks, and then order more transactions and gain more rewards, \ie, violating leadership fairness (\ssecref{subsec:issues}).
Thus, the chain quality $Q$ is defined as the expected fraction of honest blocks committed in $m$ views, given that the adversary controls a fraction of Byzantine nodes $\alpha$. We use ${B_h}_{i}$ and ${B_a}_{i}$ to denote the number of committed honest and adversarial blocks in $i$-th view, respectively. 
We have the chain quality $Q$ as:
\begin{equation} \label{eq:quality}
    Q(\alpha) =\mathbb{E} \left[ \liminf_{m \rightarrow \infty}{\frac{\sum_{i=1}^{m} {B_h}_{i}}{\sum_{i=1}^{m} {B_h}_{i} + \sum_{i=1}^{m} {B_a}_{i}}} \right].
\end{equation}
In a system without attacks (\ie, the adversary follows the protocol), $Q(\alpha) = 1-\alpha$. This is because, in a view, the probability that the adversary (resp., honest nodes) is chosen as the leader through a fair election and then produces a block is $\alpha$ (resp., $\beta$), and all blocks are committed by nodes.
Thus, for a chained BFT protocol, the optimal chain quality $Q(\alpha)$ is $1 - \alpha$. 
However, the adversary can deviate from chained BFT protocols, \ie, launching attacks, to lower the chain quality. Thus, we define the \textit{attack threshold} of chain quality as below. 
\begin{definition}[Attack threshold]\label{def:threshold}
    The attack threshold of a chained BFT protocol is the minimum fraction of Byzantine nodes controlled by the adversary to make chain quality $Q(\alpha)$ lower than $1-\alpha$. 
\end{definition}
For a chained BFT protocol, the attack threshold should be high enough (\eg, $1/3$) such that the adversary cannot decrease the chain quality. Here, $1/3$ is the maximum fraction of the Byzantine nodes assumed in chained BFT protocols. 

\subsubsection{Censorship resilience} This censorship resilience $C$ represents the fraction of committed honest blocks out of all blocks proposed by honest leaders. It quantifies the ability of honest blocks to resist censorship. The adversary may target to minimize this metric to increase its ability to manipulate transaction ordering, \ie, censorship or selective inclusion. 
Thus, we capture this by minimizing the censorship resilience $C$ calculated in $m$ views, given that the adversary controls a fraction of Byzantine nodes $\alpha$. We use ${O_h}_{i}$ to denote the honest blocks, which are eventually not included in the main chain in $i$-th view. We have:

\begin{equation} \label{eq:cencor}
C(\alpha) =  \mathbb{E} \left[ \liminf_{m \rightarrow \infty}{\frac{\sum_{i=1}^{m} {B_h}_{i}}{\sum_{i=1}^{m} {O_h}_{i} + \sum_{i=1}^{m} {B_h}_{i}}} \right].
\end{equation}
Ideally, $C(\alpha) = 1$, which implies the adversary cannot exclude any honest blocks from the main chain, \ie, optimal censorship resilience. The optimal censorship resilience is defined below. Note that if the adversary's attacks cause no loss to their blocks, the optimal attack strategies for both metrics converge. Otherwise, attacks on censorship resilience target overriding more honest blocks, while chain growth also relates to the adversarial blocks in the main chain.

\begin{definition}[Optimal censorship resilience]\label{def:censorship resilience}
    A chained BFT protocol is censorship resilient if an honest leader proposes a block after GST, the block is eventually included in the main chain.
\end{definition}

The optimal censorship resilience also implies the optimal chain quality. This is because honest blocks are unaffected, regardless of the adversary's strategy. The optimal censorship resilience is also known as reorg resilience~\cite{d2022no}. 

\section{MDP Modeling} \label{sec:MDPmodel}
In this section, we develop a systematic framework for modeling chained BFT protocols and evaluating their leadership democracy.
Our framework leverages the Markov Decision Process (MDP), a well-established mathematical model that describes the decision-making process of an agent taking actions in a specific state and transitioning to the next state, where each state transition is associated with a reward. It allows us to find utility-maximizing strategies in a stochastic environment. 
The reason for using MDPs is their ability to accurately capture the sequential and probabilistic nature of the block proposing process, allowing them to obtain the optimal adversarial strategies. Unlike game theory or discrete event simulations, MDPs efficiently handle dynamic state transitions and offer flexibility to adapt the model to different blockchain protocols and attack scenarios. 

MDP has been commonly used to analyze PoW-based Nakamoto-style consensus protocols~\cite{sapirshtein2017optimal, gervais2016security, zhang2017necessity}. MDP abstracts the chain structure and incorporates randomness in block generation. However, existing models cannot be applied to chained BFT protocols. 
The key reason is that chained BFT protocols have different rules for reaching consensus and committing blocks. {PoW protocols lack the concept of leader nodes. Instead, nodes in PoW protocols propose blocks by solving mathematical puzzles. }
The huge differences lead to different security guarantees; Nakamoto-style consensus only provides probabilistic security for blocks, whereas chained BFT protocols can ensure that a committed block will never be reverted (\ie, finality~\cite{hotstuffPODC}). All these differences require us to design new models for chained BFT protocols. 

We also observe that in chained BFT, a block has multiple statuses: being certified (\ie, forming a QC), being locked (\ie, must be extended by subsequent certified blocks), and being committed (\ie, accepted by all nodes). 
These statuses significantly increase the state space's size and the action space's complexity, making it more difficult to model the chained BFT protocols. 
Thus, we simplify the protocols and confine the adversary’s actions to build MDP modeling for various chained BFT protocols. {Such simplifications prevent the state space from exploding while still preserving the essential dynamics, making the resulting MDP solvable yet expressive enough to capture all critical decision processes and state transitions.}

\subsection{Applying MDP to Chained BFT} 
MDP is a mathematical framework for modeling decision-making situations where a single decision-maker finds the optimal strategy to maximize some designated cumulative rewards~\cite{mdp, puterman2014markov,sigaud2013markov}.
An MDP model includes four components: $\langle S, A, P, R \rangle$, where $S$ denotes all possible states, $A$ denotes a set of actions, $P$ corresponds to the transition probability function, and $R$ represents the reward for each state transition. 
Modeling chained BFT protocols using MDP then involves encoding all system states that affect the adversary's actions into the state space $S$, enumerating the adversary’s possible actions $A$, determining all possible state transitions $P$, and the associated rewards $R$ with each state transition.

\bheading{State space.}
Chained BFT protocols run in views, and each view follows a propose-vote paradigm (\ssecref{subsec:BackchainedBFT}). 
The state space $S$ then includes the block tree status at the beginning of a view, 
which contains information that affects the adversary's and the honest nodes’ decisions, such as the number of honest and adversarial blocks, and the identity of the current leader.

\bheading{Action space.} At the beginning of each view, if the adversary has a hidden block, it can choose to publish the block to fork honest blocks or for rewards. Otherwise, the hidden block is abandoned. Here, the adversary can decide that the published block reaches honest nodes before or after the subsequent honest block.
If the adversary is elected as the leader and no hidden block exists, it can attempt to create a fork on the non-locked honest blocks.

Since the adversary aims to maximize its utility, it will not follow any suboptimal strategy.
This allows us to omit certain suboptimal strategies and simplify the modeling.
For example, the adversary will not choose to initiate multiple forks simultaneously, since it does not increase the probability of overriding an honest block, yet introducing internal competition between its forks.
In addition, the adversary will attempt to override as many honest blocks as possible via a single action, rather than only overriding a subset of them.

\bheading{State transition.}
Given the current state and the adversary's action, we can determine all possible state transitions to the next states, \ie, states of the next view. 
The MDP state transition is triggered by the new mining event.
Thus, given the current state and the adversary's actions, the next state can be determined according to the newly mined block and the next leader. 
We stress that the probability distribution of state transitions depends only on the previous state and the action taken at each time step.

\bheading{Rewards.}
The reward allocation usually takes place together with the state transition.
The rewards are distributed for certain blocks if all honest nodes agree that these blocks are committed. 
Note that the reward is used to compute the utility of our two metrics. It is different from ``block rewards" used to incentivize participants in blockchains. 

{The key to MDP lies in formulating an objective function that guides the decision-making of the agent. MDP assigns rewards to different state transitions to maximize the objective function. Through the MDP model, we can systematically explore the available states for the adversary and iteratively calculate the expected objective function value of different actions. Through the iteration, the objective function converges to its maximum value, and MDP can obtain the actions that lead to the maximum objective function, thereby finding the optimal strategy.}

\subsection{MDP Design} \label{subsec:mdpdesign}
In this section, we introduce a general MDP model as a basis and will extend it to model different chained BFT protocols in \ssecref{sec:Protocolanalysis} and \ssecref{sec:countermeasure}.
We now determine the four components of MDP for modeling these protocols.

\bheading{Action space design.} In chained BFT protocols, the adversary can take the following Byzantine actions. 
\begin{packeditemize}
    \item \textbf{\texttt{Adopt}}. The adversary adopts all existing non-locked honest blocks and discards hidden adversarial blocks. If elected as the leader, it proposes a block after the adopted blocks. 

    \item \textbf{\texttt{Wait}}. The adversary waits for the leader to propose a block for the current view. If it is the leader, it will either extend its block or create a forking block to exclude the non-locked honest blocks.

    \item \textbf{\texttt{Release}}. The adversary discloses its hidden block. This action can only be taken when there exists a hidden adversarial block.
\end{packeditemize}

In chained BFT protocols, the adversary can also keep silent (\ie, proposing no block) or propose equivocating blocks during a view. However, these behaviors are not optimal for the adversary to decrease chain quality and censorship resilience. This is because they weaken the adversary's power to win forks, making honest blocks abandoned. In all cases, these actions have no benefit for the adversary on leadership democracy metrics, and the agent of the model will not choose this behavior. Therefore, we do not consider them. However, keeping silent is proven to be the key strategy to deteriorate the efficiency of chained BFT protocols~\cite{gai2021dissecting, dinh2017blockbench, shapiro2020performance, amiri2022bedrock, gramoli2023diablo}. The reason for the divergence is that the utility of the adversary has changed, which further implies that existing efficiency analysis cannot be directly adopted.  

\bheading{State space design}. The state space is a three-tuple of (\la, \lh, \leader). The \la and \lh represent the number of adversarial blocks and honest blocks that are not committed yet, respectively. The \leader is used to indicate whether the current leader is honest or an adversarial node. Specifically, we use $0$ (resp., $1$) to denote an adversarial (resp., honest) leader. 
Since each block contains a reference to the previous one, the adversary has to decide whether to release or discard its hidden block in the subsequent view. Therefore, adversarial blocks can be hidden up to one, and \la has two possible values: $\{0, 1\}$.

\bheading{State transition and rewards}.
By modeling these attack behaviors as states and actions within an MDP framework, we can quantitatively evaluate the leader democracy metrics of the chained BFT protocols under attack. By analyzing the optimal attack strategies of the adversary, we can gain more in-depth insights into the protocol's vulnerabilities and propose corresponding defense and improvement measures to enhance the protocol's security and reliability. 

For each state transition, there are some rewards for honest nodes and the adversary. Specifically, we use a tuple $(B_h, B_a, O_h)$, where $B_h$ and $B_a$ are required to record the rewards of both honest and adversarial leaders, and $B_h$ and $O_h$ are needed to calculate censorship resilience. 
A unit of reward is allocated to the leader when a block is sure to be committed. Each action of the adversary leads to a state transition, and the protocol enters the next view. 

\section{Leadership Democracy Analysis} \label{sec:Protocolanalysis}
We evaluate four representative chained BFT protocols: 2CHS, CHS, Streamlet, and FHS.
By the MDP model, we can obtain the optimal attack strategies and the corresponding leadership democracy. 
Besides, we also validate the effect of the obtained optimal strategies on chosen protocols in the open-source benchmark platform Bamboo~\cite{gai2021dissecting}. 

\subsection{Modeling Chained Protocols}
We extend the general MDP model in \ssecref{subsec:mdpdesign} to analyze and compare the leadership democracy metrics of the four chained BFT protocols when initialized with their parameters and designs.
In particular, the differences among protocols mainly affect the state transition, reward allocation, and the range of variables in the state space. 

\subsubsection{2CHS and FHS} The attack strategies and consensus rules of 2CHS and FHS are different, however, they behave the same in the MDP model. 
Specifically, in FHS, nodes send votes to the next leader, so the adversary controlling the next leader can hide the associated QC (\appref{appen:fasthotstuff}). 
In other words, the block is analogous to being overridden by the next adversarial leader in 2CHS. Therefore, these two protocols have the same state transition and reward matrices and thus have the same MDP models.
We now introduce the four components of the MDP model for 2CHS and FHS. 
The state transition and reward matrices for 2CHS and FHS are shown in \tabref{tab:state_trans_CHS}. 
\begin{table}[t]
\caption{State transition and reward matrices for 2CHS and FHS. 
The variable $\alpha$ denotes the fraction of adversarial nodes. The \texttt{release} action is feasible only when $\la>0$. The reward is a tuple of $(B_h, B_a, O_h)$.} 
\label{tab:state_trans_2CHS}
\scriptsize
\centering
\ra{1.1}
\resizebox{\linewidth}{!}{%
\begin{tabular}{@{}cccc@{}}
\toprule[1pt]
\textbf{State $\times$ Action} & \textbf{Resulting State} & \textbf{Probability} & \textbf{Reward} \\ \midrule[1pt]
\multirow{2}{*}{$(\la, \lh, 1)$, \texttt{Adopt}} & $(0, 1, 0)$ & $\alpha$ & \multirow{2}{*}{$(\lh, 0, 0)$}\\
 & $(0, 1, 1)$ & $1-\alpha$ &  \\ \midrule[1pt]
\multirow{2}{*}{$(\la, \lh, 0)$, \texttt{Adopt}} & $(1, 0, 0)$ & $\alpha$ & \multirow{2}{*}{$(\lh, 0, 0)$} \\
 & $(1, 0, 1)$ & $1-\alpha$ & \\ \midrule[1pt]
\multirow{2}{*}{$(\la, \lh, 1)$, \texttt{Wait}} & $(0, \min(\lh+1,1), 0)$ & $\alpha$ & \multirow{2}{*}{\shortstack{$(1, 0, 0)$ if $\lh=1$\\$(0, 0, 0)$ otherwise}} \\
 & $(0, \min(\lh+1,1), 1)$ & $1-\alpha$ & \\ \midrule[1pt]
\multirow{2}{*}{$(0, \lh, 0)$, \texttt{Wait}} & $(1, \lh, 0)$ & $\alpha$ & \multirow{2}{*}{$(0, 0, 0)$} \\
 & $(1, \lh, 1)$ & $1-\alpha$ & \\ \midrule[1pt]
\multirow{2}{*}{$(1, \lh, 0)$, \texttt{Wait}} & $(1, 0, 0)$ & $\alpha$ & \multirow{2}{*}{$(0, 1, \lh)$} \\
 & $(1, 0, 1)$ & $1-\alpha$ & \\ \midrule[1pt]
\multirow{2}{*}{$(1, \lh, 1)$, \texttt{Release}} & $(0, 1, 0)$ & $\alpha$ & \multirow{2}{*}{$(0, 1, \lh)$} \\
 & $(0, 1, 1)$ & $1-\alpha$ & \\ \midrule[1pt]
\multirow{2}{*}{$(1, \lh, 0)$, \texttt{Release}} & $(1, 0, 0)$ & $\alpha$ & \multirow{2}{*}{$(0, 1, \lh)$} \\
 & $(1, 0, 1)$ & $1-\alpha$ & \\ \midrule[1pt]
\end{tabular}%
}
\end{table}

\bheading{Actions.} 
The actions modeled for 2CHS and FHS are the same as those described in \ssecref{subsec:mdpdesign}.

\bheading{State space.} The state space is a three-tuple in the form of (\la, \lh, \leader), which is the same as those in \ssecref{subsec:mdpdesign}. 
Specifically, when all honest nodes have the same latest block and lock on the associated parent block, all blocks before the parent block, including the parent block itself, can be considered as being committed. (See the locking rule in \appref{appen:2chs}.) 
Therefore, given two consecutive honest blocks, only the last honest block may be forked by an adversarial block. 
Besides, in 2CHS, if an adversarial leader creates a block forking with an honest block in the previous view, the adversarial block can always win by the deterministic \texttt{proposing rule} (\appref{appen:2chs}).
{If the adversarial leader attempts to fork more than one honest block, no honest nodes will vote for this new block according to the voting rule. It is meaningless for the adversary to engage in such behavior. We only need to consider a two-depth fork. Extending the forking depth to three or more does not yield any additional attack advantage and leaves our fairness bounds unchanged.}
Therefore, \lh can be simplified to have two values: $\{0, 1\}$. 
The value range of $\la$ and \leader is the same as that in the general model. 

\bheading{State transition and reward allocation.} 
Given states and the adversary's action, we can decide $\la$ and $\lh$ for the next state and the associated transition rewards. 

\begin{packeditemize}
    \item \textbf{\texttt{Adopt}}. The adversary accepts honest blocks, and so the system returns to its initial state, where the honest nodes and the adversary hold the same chain.
    Thus, both \la and \lh become 0, and honest nodes receive reward $B_h = \lh$. Then, the current honest (resp., adversarial) leader proposes one block, \ie, \lh (resp., \la) increases by 1.  

    \item \textbf{\texttt{Wait}}. When the current leader is honest, \lh increases by 1. Since the maximum value of \lh is 1, so \lh is updated to $\min($\lh$+1, 1)$. Specifically, when $\lh=1$, the first honest block becomes committed and so $B_h = 1$. 
    
    When the current leader is an adversary, and there is no hidden adversarial block, it will launch a forking attack and cause an increment of $1$ in \la. Otherwise, it will extend the adversarial block, which reveals the previously hidden adversarial block, and \lh will be reset to 0. Under this case, the adversary will win one block reward, \ie, $B_a = 1$, and honest blocks will be abandoned $O_h = \lh$.  

    \item \textbf{\texttt{Release}}. Since the hidden block is published, the previous unrewarded honest blocks (if any) will be overridden, and the adversarial block will be rewarded, so both \la and \lh are reset to 0. Thus, the adversary will win one block reward, \ie, $B_a = 1$, and honest blocks will be abandoned $O_h = \lh$. 
    After that, the current leader generates a block, and so \la or \lh increases by 1.
\end{packeditemize}

There are two possible results for the next leader $L$, being adversarial (\ie, $0$) and honest (\ie, $1$). Specifically, the probability for $L$ equals to $0$ (resp., $1$) is $\alpha$ (resp., $\beta$). (See leader election in \ssecref{sec:model}.)

\subsubsection{CHS} 
Due to the three-chain structure, CHS has a different lock rule, where a node will lock on the grandparent block given three consecutive blocks. Furthermore, \lh can be simplified to have three values, \ie, $\{0, 1, 2\}$. 
Besides, we simplify the forking case, in which given two uncommitted honest blocks, the adversary will fork them at once.
In other words, we do not consider the forking attack of only one honest block. This simplification will not affect the results since, in this case, the adversary can always make the two honest blocks not included in the main chain by the deterministic consensus rule. Thus, making one honest block abandoned in this case is not the optimal strategy for the attacker to maximize its utility (\ie, lowering chain quality and censorship resilience).  
If there are two non-locked blocks: one is honest, and the other is adversarial. When the honest block comes first, it indicates that it has been adopted, and at this point $\lh=0$; Otherwise, the adversary can fork this honest block.
By this simplification, the same state space, actions, state transition, and reward allocation of CHS are the same as those in 2CHS, except for the possible values of \lh. 
Due to space constraints, we provide the state distribution and reward allocation of CHS in \tabref{tab:state_trans_CHS} at \appref{appen:mdpmodel}.

\subsubsection{Streamlet} Streamlet differs from the other three chained BFT protocols in two aspects. First, it follows the longest certified chain rule, which means honest nodes only vote for blocks that extend the longest certified chain. 
Second, in Streamlet, there is no view-change mechanism for the leader to synchronize the highest certified blocks. Thus, an honest leader may propose a block not voted by other honest nodes.
Due to these differences, the adversary can strategically publish its hidden certified block to make no honest nodes vote for subsequent honest blocks. (See more details of the forking example in \appref{appen:Streamlet}.)
Taking these differences into consideration, we introduce the MDP model of Streamlet. Due to space constraints, the associated state distribution and reward allocation are provided in \tabref{tab:state_trans_CHS} at \appref{appen:mdpmodel}.

\bheading{Actions.} We introduce an additional \texttt{withhold} action to indicate the above attack strategy. 
This action means that the adversary publishes its hidden certified block after that the current honest leader proposes a block. 

\bheading{State space.} The state space of Streamlet is the same as that in the general model. 

\bheading{State transition and reward allocation.} 
We can determine the next state and the corresponding rewards by the current states and the adversary's action.
\begin{packeditemize}
    \item \textbf{\texttt{Adopt}}. The state transition and associated rewards are the same as 2CHS.

    \item \textbf{\texttt{Wait}}. When the current leader is an adversarial node, $\la=0$ and $\lh > 0$, since there is no forking attack, the adversary's attempt to override the honest blocks will fail, so \la remains unchanged at 0. 
    If the current leader is adversarial and there is a hidden adversarial block, the adversary receives a reward $B_a=1$, and the unrewarded honest blocks will be rewarded with $B_h=\lh$.

    \item \textbf{\texttt{release}}. The state transition is the same as 2CHS. The adversarial block and the previous honest blocks (if any) will be rewarded, so $B_h$ receives a reward of \lh and $B_a$ receives a reward of 1.
    
    \item \textbf{\texttt{Withhold}}. With the exposure of the hidden adversarial block, the state will reset. If the current leader is honest, the new proposed block will be invalidated by adversarial nodes, so \lh remains 0. And if the leader is an adversary, \la becomes 1.
    The adversary receives a reward of $B_a=1$, and honest nodes receive $B_h=\lh$.
    When the current leader is honest, and a hidden adversarial block exists, the proposed honest block is abandoned, and a reward loss occurs. At this point, $O_h$ receives a reward of 1.
    
\end{packeditemize}

\begin{figure}[t]
    \centering
    \includegraphics[width=0.9\linewidth]{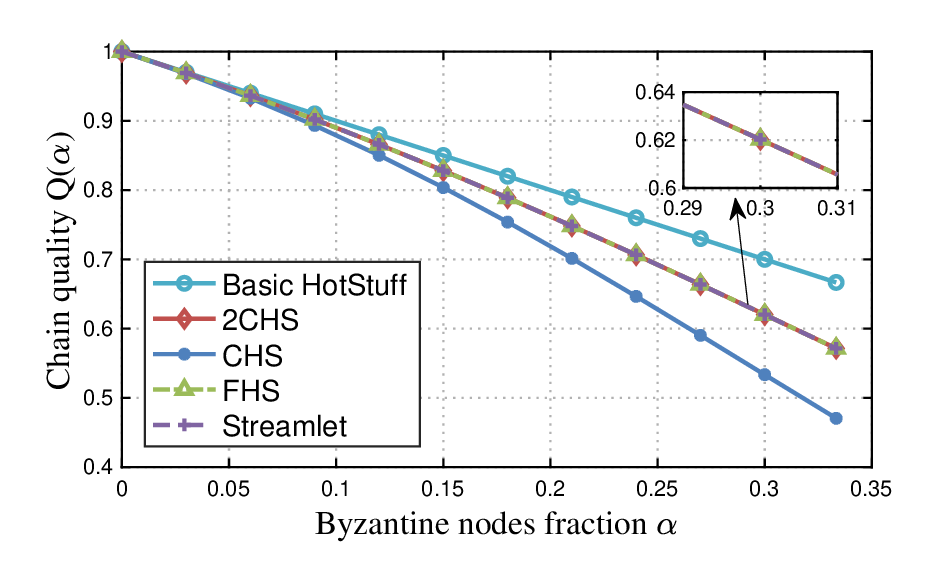}
    \caption{The chain quality of 2CHS, CHS, FHS, and Streamlet. The $Q(\alpha)$ values of 2CHS, FHS, and Streamlet overlap with each other.}
    \label{fig:chainQuality}
\end{figure}

\subsection{Evaluation Results} \label{subsec:findindOne}

\subsubsection{Modeling Objective Functions and MDP Settings}
Our modeling is not a standard MDP, because the objective function $Q(\alpha)$ or $C(\alpha)$
is non-linear. To apply standard MDP techniques, we follow the procedure developed by Sapirshtein et al. \cite{sapirshtein2017optimal} that transforms a non-standard MDP into a family of standard MDPs.

Here, we briefly describe the procedure with $Q(\alpha)$ as a particular example and we refer our readers to \cite{sapirshtein2017optimal} for more details. First, we define a new objective function $Q'(\alpha)$ as:
\begin{equation} 
    Q'(\alpha) =\mathbb{E} \left[ \liminf_{m \rightarrow \infty}{\frac{\sum_{i=1}^{m} {B_a}_{i}}{\sum_{i=1}^{m} {B_a}_{i} + \sum_{i=1}^{m} {B_h}_{i}}} \right].
\end{equation}
The adversary aims to maximize $Q'(\alpha)$.
Suppose that the value of $Q'(\alpha)$ is $\rho$. Define, for any $\rho \in [0, 1]$, the
transformation function $w_{\rho}: \mathbb{N}^2 \to \mathbb{R}$ as:
\[
w_{\rho}(B_a, B_h) = (1 - \rho) B_a - \rho B_h,
\]
where $B_a$ and $B_h$ are the reward of the adversary
and the honest nodes, respectively.
This gives rise to a family of standard MDPs $\langle S, A, P, w_{\rho}(R) \rangle$ (parameterized by $\rho$) that share the same state space, action, and transition matrix as the original problem but the reward matrix is determined by $w_{\rho}$.

Let $v_\rho^{\pi}$ be the expected value of $\langle S, A, P, w_{\rho}(R) \rangle$ under policy $\pi$, \ie,
\[
v_\rho^{\pi} = \mathbb{E}\left[ \liminf_{m \rightarrow \infty} \frac{1}{m} \sum_{i=1}^{m} w_{\rho}\left({B_a}_{i}(\pi), {B_h}_{i}(\pi)\right) \right].
\]
Let $v_\rho^{*}$ be the expected value under the optimal policy, \ie, $v_\rho^{*} = \max_\pi \left\{ v_\rho^{\pi} \right\}$. We have the following properties according to \cite{sapirshtein2017optimal}:
\begin{packeditemize}
    \item $v_\rho^{*}$ is monotonically decreasing in $\rho$ for $\rho \in [0, 1]$.
    \item If $v_\rho^{*} = 0$ for some $\rho \in [0, 1]$, then an optimal policy $\pi^{*}$ for $\langle S, A, P, w_{\rho}(R) \rangle$ also maximizes $Q'(\alpha)$ with its optimal value given by $\rho$.
\end{packeditemize}

Note that $v_0^{*} > 0$ and $v_1^{*} < 0$. The first property implies that we can use a standard binary search to find the value of $\rho$ under which $v_\rho^{*} = 0$. Let us denote this value by $\bar{\rho}$.
The second property says that $\bar{\rho}$ is the maximum possible $Q'(\alpha)$, which equals to saying that $1 - \bar{\rho}$ is the minimum possible $Q(\alpha)$.
In this way, we compute the minimum possible $Q(\alpha)$ with $\alpha$ between $0$ and $0.33$ (\ie, 1/3 the maximum fraction of Byzantine nodes in the system) with interval $0.03$ and predefined maximum error of $10^{-4}$.

\subsubsection{Evaluation Results}
In the ideal case where no adversarial behaviors exist, the protocol achieves optimal chain quality (\ie, $Q(\alpha) = 1-\alpha$) and censorship resilience (\ie, $C(\alpha) = 1$). We thus use them as the optimal values for comparing with the evaluation results. 
\figref{fig:chainQuality} shows the chain quality of the four protocols under different fractions $\alpha$. Some findings are listed as follows. 

\begin{finding}
The chain quality $Q(\alpha)$ of the four protocols is lower than the optimal values; the chain quality decreases with increasing $\alpha$.
\end{finding}

The decrease in chain quality across all $\alpha$ indicates that the \textit{attack thresholds} (defined in \ssecref{subsec:metrics}) of these protocols are zero. In other words, the adversary with an arbitrarily small $\alpha$ ($\alpha > 0$) would deviate from protocols to lower the chain quality. The root cause is that the adversary can achieve riskless attacks due to the deterministic rules. Specifically, the adversary can always launch forking attacks with honest blocks and ensure that its block is included in the main chain. 
Besides, more adversarial nodes lead to worse chain quality, as adversarial nodes have more opportunities to create forking blocks to substitute honest blocks from the main chain. 

\begin{figure}[t]
    \centering
    \includegraphics[width=0.9\linewidth]{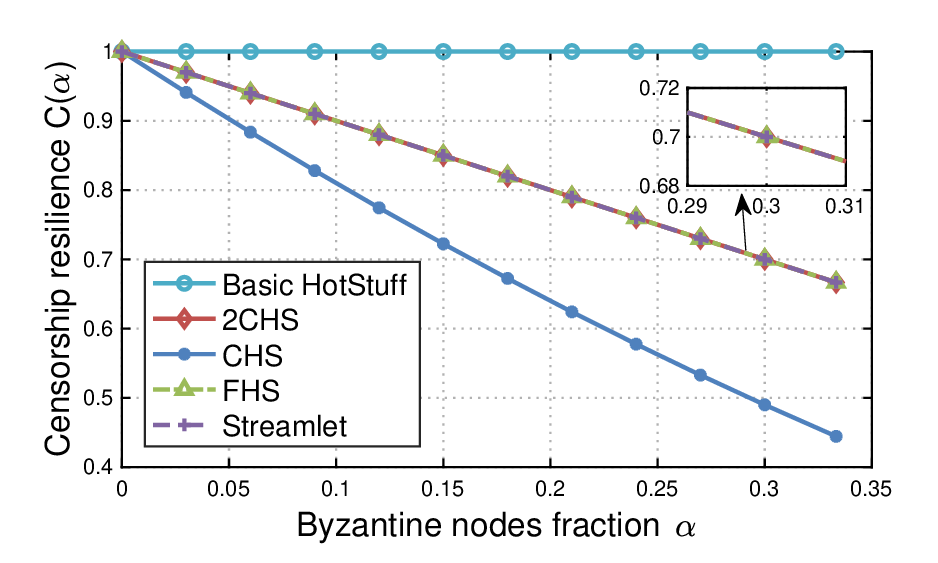}
    \caption{The censorship resilience of 2CHS, CHS, FHS, and Streamlet. The $C(\alpha)$ values of 2CHS, FHS, and Streamlet overlap with each other.}
    \label{fig:censorship}
\end{figure}

\begin{finding}
The chain quality $Q(\alpha)$ of 2CHS, FHS, and Streamlet are the same, and CHS has the lowest chain quality among the four protocols. 
\end{finding}

The lines of 2CHS, FHS, and Streamlet in the figure overlap, indicating that they have the same chain quality under different $\alpha$. However, the reasons behind the deterministic attack opportunities for each protocol are different.
CHS adopts a three-chain structure and extends the block with the highest QC when proposing, which allows the adversary to override two honest blocks in CHS, while other protocols can override only one block.

\figref{fig:censorship} shows the censorship resilience of the four chained BFT protocols under different $\alpha$. Our findings are as follows. 

\begin{finding}
    All four protocols are susceptible to censorship 
 attacks, and CHS exhibits worse censorship resilience compared to the other protocols.
\end{finding}

The existence of forking attacks leads to excluding honest blocks. For 2CHS, CHS, and FHS, it is a posterior forking attack, while for Streamlet, it is a preemptive forking attack. Meanwhile, the attacks on CHS involve more honest blocks, resulting in greater losses. For the other three protocols, one adversarial block can censor one honest block, while the adversary in CHS can censor up to two blocks.

\subsection{What Goes Wrong with Chaining?}
Our analysis results highlight a noteworthy challenge: none of the examined protocols—2CHS, CHS, FHS, and Streamlet—attains a satisfying  
chain quality and censorship resilience. This deficiency mandates a closer examination of the protocols' vulnerabilities, leading to two important insights.

\begin{packeditemize}
    \item \textbf{Leader rotation is not enough for leadership democracy.} 
    Some protocol designs lack consideration for the impact on the leadership democracy.
    The forking attacks in 2CHS and CHS allow the adversary to consistently override non-locked honest blocks, undermining both chain quality and censorship resilience.
    FHS eliminates direct forking attacks by collecting the proof of the latest QC before the block proposal; however, it overlooks the transference of QC formation rights to the succeeding leader, leaving a residual vulnerability. Streamlet conceals adversarial blocks to prevent the new proposed block from receiving enough votes to be valid. 
    The chaining structure of these four protocols allows the adversary to substitute honest blocks. 
    So it is important to consider the impact when designing a protocol and prepare a better evaluation framework in advance. 

    \item \textbf{The adversary always wins in forking attacks.} 
    Although the adversary adopts different attack strategies to create forks with uncommitted honest blocks among protocols, it can always make these honest blocks not included in the main chain. 
    Thus, even when controlling a small number of Byzantine nodes, the adversary can profit from launching attacks. This riskless nature of attacks underscores the urgent need for enhanced security measures to protect these protocols against adversarial exploits.
\end{packeditemize}

\subsection{Experiments}
With the framework, we can obtain the optimal attack strategies and the theoretical performance of leadership democracy for chained BFT protocols.
We then validate the effects of these optimal strategies through experiments conducted on the open-source benchmark platform, Bamboo~\cite{gai2021dissecting}. {We mainly focus on analyzing without honest leader timeouts in the section below, while considering the timeouts in \appref{append:gamma-expe}.}

\bheading{Implementation and system setup.} We extend Bamboo~\cite{gai2021dissecting} developed in the Go language to implement the chained BFT protocols. We integrate strategies for forking attacks by modifying the proposing and voting actions of the adversary. Our modifications amount to roughly 150 lines of code, facilitating the implementation of 2CHS, CHS, FHS, and Streamlet. 

The experimental setup includes 4 servers, each equipped with an 8-core CPU, 16GB RAM with the operating system of Ubuntu Server 22.04. 
To simulate a realistic network environment, we design a network topology capable of supporting up to 60 nodes (with each server handling 15 nodes) and allowing the inclusion of up to 18 Byzantine nodes to evaluate the system's leadership democracy under attack. 
We conduct 6 separate runs for each group of experiments of 4000 views and obtained the average results.

\begin{figure}[t]
    \centering
    \includegraphics[width=0.9\linewidth]{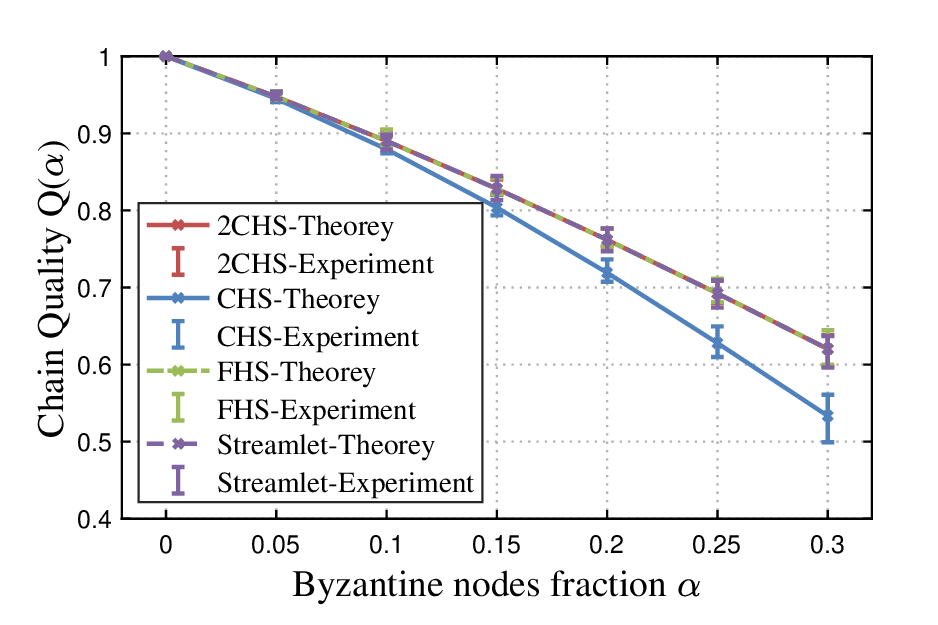}
    \caption{Comparison between experimental results (denoted by the error bar) and theoretical  results (denoted by the x symbol) of chain quality of 2CHS, CHS, FHS, and Streamlet.}
    \label{fig:quality_experiment}
\end{figure}

\bheading{Experimental results of chain quality.}
\figref{fig:quality_experiment} shows the experimental and theoretical results of chain quality under different Byzantine node fractions $\alpha$ of the four protocols. 
As we can see, the theoretical values fall within the experimental range.
The \metricOne of CHS, 2CHS, FHS, and Streamlet gets worse as $\alpha$ increases. 
When $\alpha\!=\!0.3$, \metricOne of 2CHS drops to $62\%$ of that when $\alpha\!=\!0$. 
CHS has the worst chain quality among the four protocols, while 2CHS, FHS, and Streamlet show similar chain quality.

\begin{figure}[t]
    \centering
    \includegraphics[width=0.9\linewidth]{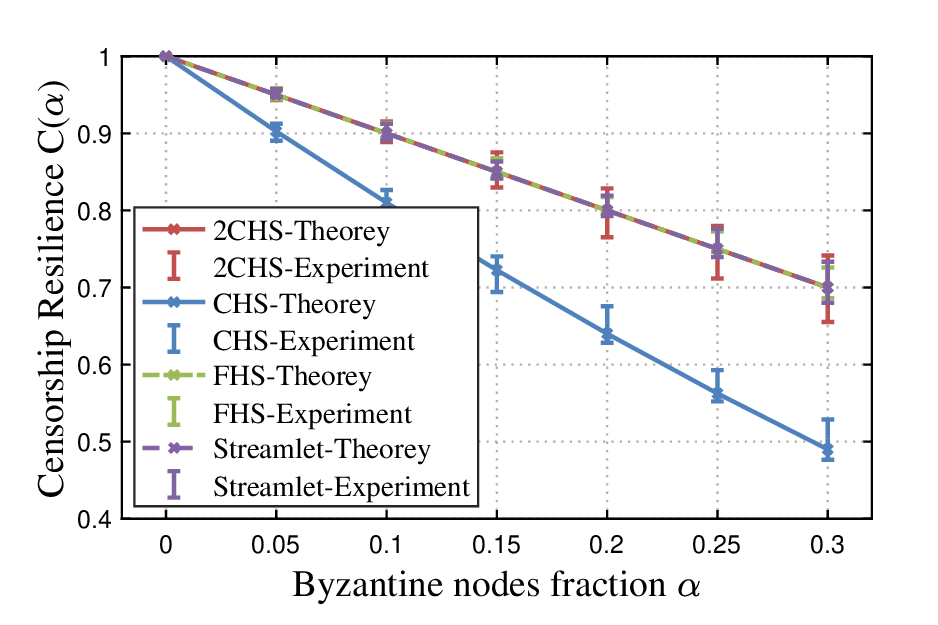}
    \caption{Comparison between experimental results (denoted by the error bar) and theoretical  results (denoted by the x symbol) of censorship resilience of 2CHS, CHS, FHS, and Streamlet.}
    \label{fig:censorship_experiment}
\end{figure}

\bheading{Experimental results of censorship resilience.}
\figref{fig:censorship_experiment} shows the \metricTwo results under different $\alpha$. The experimental results closely match the theoretical values. The results also show CHS has the worst censorship resilience among these protocols.
As $\alpha$ increases from 0 to 0.3, \metricTwo of FHS decreases by $30\%$, and CHS decreases by $49\%$.

{The experimental results fluctuate within a certain range. This is because the effect of the attack is uncertain since the leader is elected randomly. Measurement errors grow as $\alpha$ increases, as chances for the adversary to launch attacks increase with $\alpha$. Take CHS as an example, the relative error at $\alpha=0.05$ is about $0.9\%$, while the relative error at $\alpha=0.3$ is about $6\%$.} The maximum relative error is $6\%$, indicating that our experimental results are stable. The experiments prove the validity and robustness of our framework in evaluating chain quality and censorship resilience (\ssecref{subsec:findindOne}).

\section{Countermeasures and Analysis}\label{sec:countermeasure}
In this section, we propose countermeasures against attacks on the leadership democracy for these chained BFT protocols and analyze their effectiveness within our framework. Our countermeasures should not modify \texttt{voting and committing rules}, ensuring that the safety and liveness properties of chained BFT protocols are not affected. Thus, we do not need to prove the protocols' security again. Besides, our countermeasures should not compromise the inherent characteristics of these protocols, such as linear message complexity and responsiveness in CHS, as well as preserving two-chain latency in FHS.

Guided by the above principles, we propose countermeasures to mitigate the vulnerabilities and weaknesses of each protocol found in the previous analysis. 
\begin{packeditemize}
    \item \textbf{FHS-C with broadcasting QCs (\S\ref{subsec:broadcasting-qcs})}. The weakness of FHS lies in the \textit{voting pattern} of DV; an adversarial leader can hide QCs of honest blocks in the previous view. Therefore, we replace DV with LBV (c.f., \ssecref{subsub:components}). 

    \item \textbf{2CHS-C/CHS-C with random proposing rule and broadcasting QCs (\S\ref{subsec:random-proposing-rule})}. 2CHS and CHS have two weaknesses. 
    First, same as FHS, the \textit{voting pattern} DV in 2CHS and CHS is replaced by LBV\footnote{The original version of CHS adopts LRV~\cite{hotstuffPODC} and later uses $DV$~\cite{hotstuffArxiv}.}.
    Second, the deterministic rule allows the adversary to always win in the posterior forking attacks. We introduce randomness to the \texttt{proposing rule} such that attacks do not always succeed.
\end{packeditemize}

Here, we use LBV rather than BV because the former does not change the message complexity (see our principles of countermeasures). However, LBV requires one additional round for broadcast compared with DV, increasing the latency. 

\subsection{Broadcasting QCs in FHS}
\label{subsec:broadcasting-qcs}

\subsubsection{Broadcating QCs} 
We observe that posterior forking attacks of FHS are caused by the voting pattern DV, in which an adversarial leader can hide the QC of the previous honest block and then invalidate it. Circumventing these attacks requires a design for preventing attackers from hiding QC.

To this end, we propose to replace the voting pattern DV with LBV in FHS, i.e., the QC is collected and broadcast by the current leader rather than the next leader.
Then, the honest leader will collect votes for its block and then broadcast the associated QC to all nodes in the current view, and the adversarial leader cannot hide the QC of other blocks.
Since all honest nodes know the certified block, the next leader must extend this block according to the \texttt{voting rule} of FHS, circumventing the posterior forking attacks. {This modification in the voting pattern will introduce some acceptable performance overhead, referring to the relevant experiments and analysis in \appref{append:fhs-c}.}

\subsubsection{FHS-C analysis} We first prove that FHS-C has optimal censorship resilience, as per Definition~\ref{def:censorship resilience}. In the following analysis, we consider the worst case that there are $f$ Byzantine nodes (\ie, $t =f$).  

\begin{theorem}\label{lemma:fhs-c-censorship-resililence}
    FHS-C has optimal censorship resilience regardless of the adversary's strategy.
\end{theorem}
\begin{IEEEproof}
    For the sake of contradiction, assuming the adversary can break the optimal censorship resilience.
    That is, the adversary can override a block $B_v$ proposed by an honest node in view $v$ by using a conflicting block $B_{v'}$  in view $v' \neq v$ at the same height.
    In other words, $B_{v'}$ rather than $B_v$ is committed at the height. 
    Let $v'$ be the view number where an adversarial node is elected as leader and proposes this conflicting block.
    Then, we have the following cases.
    \begin{packeditemize}
        \item $v' \leq v-2$: By FHS's two-chain \texttt{committing rule}, $B_{v'}$ is committed by all honest nodes at view $v$. This means $B_{v'}$ is in the prefix of $B_v$ in every honest node's view. However, $B_v$ is at the same height as $B_{v'}$, contradicting the safety property where the chain ended with $B_{v'}$ should be a prefix of the chain ended with $B_v$ for every honest node.
        
        \item $v' = v-1$: Since $B_v$ is conflicting with $B_{v'}$, the honest leader does not observe $B_{v'}$ at view $v$. That is, the adversary withholds $B_{v'}$ at view $v'$, waits for $B_v$ to be proposed at view $v$, then publishes $B_{v'}$ to make $B_{v'}$ committed, \ie, \textit{preemptive forking attack}. However, upon receiving $B_v$, all honest nodes will vote for $B_v$ and update the highest QC to $B_v.QC$. Then, by the LBV design introduced in FHS-C, the leader at view $v+1$ collects the highest QC as $B_v.QC$, proposes a block $B_{v+1}$ and locks $B_v$. This contradicts the safety property where $B_v$ is supposed to be the same as $B_{v'}$.
        
        \item $v' = v+1$: Since $B_v$ is conflicting with $B_{v'}$, the adversarial leader at view $v'$ proposes block $B_{v'}$ conflicting with $B_v$, \ie, \textit{posterior forking attack}. This requires $B_{v'}$ to use $B_{v-1}.QC$ as the highest QC, and requires the adversary to make $\geq f+1$ honest nodes to support $B_{v'}$. However, all honest nodes have voted $B_v$ at view $v$ and thus considered $B_v.QC$ as the highest QC, which contradicts the $n=3f+1$ assumption.
        
        \item $v' \geq v+2$: By FHS's two-chain \texttt{committing rule}, $B_v$ is committed by all honest nodes at view $v'$. This means $B_v$ is in the prefix of $B_{v'}$ in every honest node's view. However, $B_{v'}$ is at the same height as $B_v$, contradicting the safety property where the chain ended with $B_v$ should be a prefix of the chain ended with $B_{v'}$ for every honest node.
    \end{packeditemize}
    Therefore, to break optimal censorship resilience, the adversary either needs to break the $n=3f+1$ assumption or break the safety property of FHS, which is impossible.
\end{IEEEproof}

\begin{corollary} \label{theo:fhs-c}
    FHS-C has optimal chain quality regardless of the adversary's strategy.  
\end{corollary}

\begin{IEEEproof}
    By Theorem~\ref{lemma:fhs-c-censorship-resililence}, every honest block can be committed and included in the main chain regardless of the adversary's strategy. Besides, the probability that the adversary (resp., honest nodes) is chosen as the leader through a fair election and then produces a block is $\alpha$ (resp., $\beta$).  
    Thus, the minimum chain quality is $\beta = 1-\alpha$, which is optimal.
\end{IEEEproof}

\subsection{Random Proposing Rule in 2CHS and CHS}
\label{subsec:random-proposing-rule}

\begin{figure}[t] 
    \centering
    \includegraphics[width=0.95\linewidth]{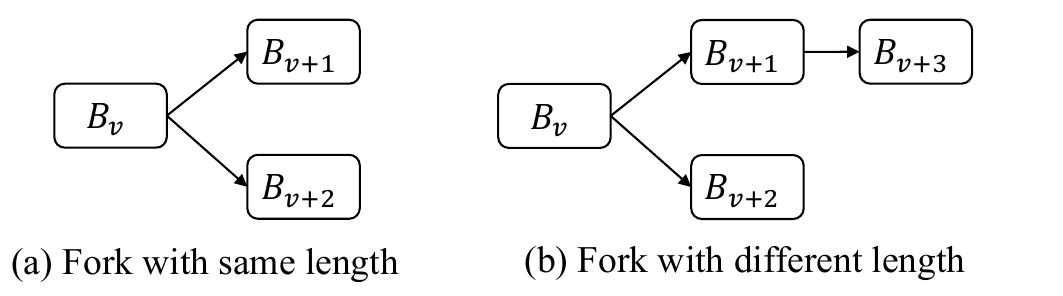}
    \caption{Possible forks in 2CHS-C and CHS-C. An honest leader first shortlists the longest forks in case (b) and follows a uniform tie-breaking rule when they are equal, as shown in case (a).}
    \label{fig:counter-random}
\end{figure}

\begin{table*}[t]
\scriptsize
\ra{1.2}
\caption{The chain quality and censorship resilience evaluation of 2CHS, 2CHS-C, CHS, and CHS-C with different $\alpha$. We have specifically marked the improved attack thresholds, $\alpha=0.285$, of chain quality.}
\label{tab:countermeasure-eval}
\resizebox{\textwidth}{!}{%
\begin{tabular}{@{}c|ccccc|cccccc@{}}
\toprule[1pt]
  \multirow{2}{*}{Protocols} & \multicolumn{5}{c|}{Chain quality} &  \multicolumn{5}{c}{Censorship resilience} \\ &
  $\alpha=0.20$ &
  {\bf $\alpha=\textbf{0.285}$} &
  $\alpha=0.286$ &
  $\alpha=0.30$ &
  $\alpha=1/3$ &
  $\alpha=0.15$ &
  $\alpha=0.20$ &
  $\alpha=0.25$ &
  $\alpha=0.30$ &
  $\alpha=1/3$ \\ \midrule[1pt]
 2CHS &0.762 & 0.642 & 0.641 & 0.620 & 0.571 &
 0.850 & 0.800 & 0.750 & 0.700 & 0.667
 \\
 2CHS-C & 0.800 & \textbf{0.715} & 0.713 & 0.693 & 0.643 &
 0.901 & 0.856 & 0.806 & 0.750 & 0.710
 \\
 \hline
 CHS & 0.719 & 0.562 & 0.560 & 0.533 & 0.471 &
 0.722 & 0.640 & 0.563 & 0.490 & 0.444
 \\
 CHS-C & 0.800 & \textbf{0.715} & 0.713 & 0.692 & 0.640 &
 0.899 & 0.853 & 0.799 & 0.737 & 0.693
 \\
\bottomrule[1pt]
\end{tabular}%
}
\end{table*}

\subsubsection{Random proposing rule}
We observe that in both 2CHS and CHS, the adversary can always succeed in posterior forking attacks. Circumventing posterior forking attacks in these protocols will require breaking changes to \texttt{voting} or \texttt{committing} rules. This is not aligned with our countermeasure principles, and we will leave such protocol designs for future work.
We instead consider a weaker level of security guarantees, where posterior forking attacks are not always successful.
This improves the expected leader democracy metrics when the attacks fail, and the optimal attack strategies will advise the adversary not to attack when the expected reward is negligible or zero.
Thus, the protocols will achieve better leader democracy metrics, although not optimal.

To this end, we introduce randomness to the \texttt{proposing rule}, i.e., a leader can randomly choose a forking branch to extend (rather than deterministically picking one). 
As depicted in \figref{fig:counter-random}, suppose the adversarial leader of view $v+2$ intentionally creates a fork. In 2CHS and CHS, the next honest leader will extend the block with the highest QC (\ie, $B_{v+2}$) according to the deterministic \texttt{proposing rule}.
In contrast, if using the random \texttt{proposing rule}, when an honest leader observes a fork at the same length as the canonical chain, it will randomly choose a fork to extend. If the two forks have different lengths as shown in \figref{fig:counter-random}(b), the honest leader will select the longer one.
Consequently, the adversary cannot always succeed in forking attacks, thereby improving the leader democracy metrics.
{We apply uniform randomness to all competing forks of equal length. Because any weighting based on fork length or view number would reintroduce exploitable bias,  leading to preemptive or posterior forking attacks.}

\subsubsection{MDP modeling}
\begin{figure}[t]
    \centering
    \includegraphics[width=0.8\linewidth]{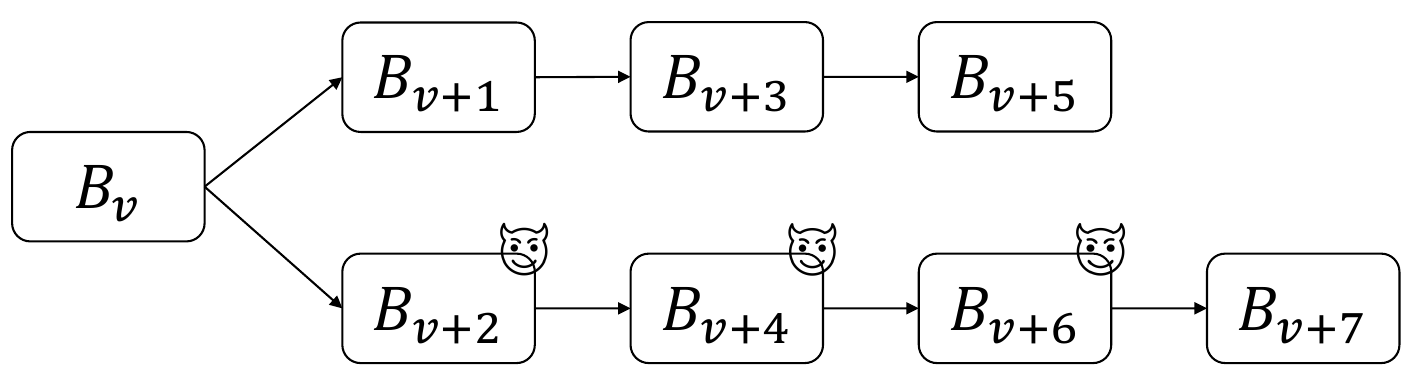}
    \caption{Forking case in 2CHS-C. The adversarial blocks are denoted using the devil icon, while the others are honest blocks.}
    \label{fig:2CHSc-fork}
    \vspace{-0.5cm}
\end{figure}

Taking 2CHS as an example, before adding randomness, according to the \texttt{voting rule}, the view number of the parent block is not less than the locked view, so there is only one manipulable honest block. However, after adding randomness, the situation in \figref{fig:2CHSc-fork} also becomes possible. Adversarial Blocks are denoted using the devil icon. After the generation of $B_{v+3}.QC$, at least $f+1$ honest nodes are locked on $B_{v+1}$. However, the adversarial block $B_{v+4}$ has a parent block with a view number of $v+2$, which is greater than $v+1$. Therefore, honest nodes locked on $B_{v+1}$ will still vote for $B_{v+4}$. The length of the fork is not necessarily 1 anymore, thus the values for \la and \lh have no fixed range. 
This situation will be broken when two consecutive blocks appear. As shown in the figure, {with blocks $B_{v+6}$ and $B_{v+7}$ forming a consecutive structure, the forking branch located below will win the fork.}
Even if the adversary tries to propose $B_{v+8}$ after $B_{v+5}$, it will not receive enough votes because its parent block view $v+5$ is smaller than $v+6$. Therefore, the first and its preceding blocks will be committed once two consecutive blocks appear.
Suppose the adversarial block wants to win in the fork. In that case, it can either rely on good luck to be extended by the next honest leader or expect the next leader to be adversarial and propose a consecutive adversarial block.
Due to space constraints, the state transition and reward matrices for 2CHS-C and CHS-C are shown in \appref{appen:randomness}.

\bheading{2CHS-C modeling.}
To represent this consecutive honest chain structure, we extend the previous MDP model and introduce a new attribute called \cnt, which indicates the number of consecutive honest blocks in the chain. The possible values of \cnt are 0 and 1, and the length of \la and \lh is no longer limited. 
We also introduce $\gamma$ and $1-\gamma$ to denote the probability that an honest leader chooses an honest branch in a fork and that it chooses the adversarial branch, respectively. 
Other attributes and actions remain the same. 
The state transition and reward allocation are determined by the current state and actions taken by the adversary.

\begin{packeditemize}
    \item \textbf{\texttt{Adopt}}. If the current leader is honest, \cnt will become 1. If the leader is an adversary, \cnt will reset to 0. The honest nodes receive a reward of $B_h=\lh$.
    
    \item \textbf{\texttt{Wait}}. There are two situations for both honest and adversarial leaders. 
    
    \bheading{Leader is honest.} 1) If $\lh=0$ or if $\lh>0, \la\le\lh$, the honest leader will naturally produce a block, and both \lh and \cnt will increase by 1. 2) If $\la>\lh>0$, it is equivalent to $\la=\lh+1$ because the honest branch will be discarded once the adversarial branch is longer. In this case, the honest leader observes two branches of equal length and will randomly choose one to extend. If the honest branch is chosen, both \lh and \cnt will increase by 1. Otherwise, a new branch will be formed with the unpublished hidden block, and \la, \lh, and \cnt will all become 1. And if the adversarial branch is chosen, the adversary will receive a reward of $B_a=\la-1$, while the honest leader will lose a reward of $O_h=\lh$.
    
    \bheading{Leader is adversarial.} 1) If $\lh=0$ or $la>lh>0$, the adversary can form a longer branch to win, which leads to a reset in the state transition. The adversary will receive a reward of $B_a=\la$, while the honest nodes will lose $O_h=\lh$ rewards. 2) If $\lh>0$ and $\la<=\lh$, the adversary chooses to extend their branch and wait for an opportunity to invalidate honest blocks. In this case, \la increases by one, and \cnt is reset to 0.
    
    \item \textbf{\texttt{Release}}. There are three cases for an honest leader and two cases for an adversarial leader. 
    
    \bheading{Leader is honest.} 1) If $\lh=0$ or if $\la>\lh>0$, the adversarial block is revealed, and the honest leader produces a new block after it, with \la being reset to 0, \lh and \cnt both set to 1. There will be $B_a=\la$ and $O_h=\lh$. 2) If $\la=\lh>0$, the disclosure of the adversarial block results in the appearance of two equal-length branches. In this case, there is a probability of $\gamma$ for extending the honest block, and both \lh and \cnt will increase by 1. If the adversarial block is extended with a probability of $1-\gamma$, \la will be set to 0, and both \lh and \cnt will be set to 1. The adversary will receive a reward of $B_a=\la$, while the honest leader will lose a reward of $O_h=\lh$. 3) If $\lh>0$ and $\la<\lh$, the honest leader will extend the honest block because it is the longest branch, and both \lh and \cnt will increase by 1.

    \bheading{Leader is adversarial.} The state transition and reward allocation are the same as \texttt{wait} action for the adversarial leader.
\end{packeditemize}

It is important to note that when \cnt accumulates to 2, it indicates that the previous honest block needs to be locked, and there will be a reward of $B_h=\lh$ for the honest leaders. Then \cnt will revert to 1, \la will be set to 0, and \lh becomes 1. This locking mechanism ensures that the longest honest chain is preserved and that the adversarial branch will be discarded, enhancing the security of the 2CHS-C protocol.

\bheading{CHS-C modeling.} 
We now model CHS-C using MDP. 
The main difference is that when there are three consecutive blocks, the first block is locked. 
Thus, the MDP modeling for CHS-C differs from that of 2CHS-C in terms of the \cnt attribute, which can take values of $\{0, 1, 2\}$. When \cnt accumulates to 3, a reward allocation occurs, with $B_h = \lh-1$. Afterward, \lh becomes 2, \la becomes 0, and \cnt becomes 2.

\subsubsection{Evaluation results} \label{subsec:findindTwo} We use the same MDP setup as analysis in \ssecref{sec:Protocolanalysis}.
We set $\gamma$ to $0.5$ since honest leaders randomly choose branches in a fork. \tabref{tab:countermeasure-eval} shows the chain quality and censorship resilience of 2CHS-C and CHS-C. Here are some findings. 

\begin{finding} 
For chain quality, 2CHS-C and CHS-C can effectively increase the attack threshold.
\end{finding}

The results indicate that the chain quality of original protocols differs from the baseline when $\alpha$ exceeds 0. It is not until $\alpha$ reaches 0.286 that 2CHS-C and CHS-C deviate from the optimal value. This suggests that the countermeasures effectively increase the attack threshold. When the proportion of the adversary is lower than the threshold, they tend to refrain from launching an attack to avoid losing their rewards.

\begin{finding}
The thresholds for 2CHS-C and CHS-C are the same. Overall, countermeasures are more effective on CHS compared to 2CHS in chain quality.
\end{finding}

When $\alpha = 0.286$, 2CHS-C and CHS-C have the same chain quality value. The optimal strategies in CHS-C show that the adversary in CHS-C tends to attack one honest block when $\alpha$ is not large enough. By contrast, in CHS, the optimal strategy for the adversary is always forking two honest blocks. The difference is because of the random \texttt{proposing rule}, by which the adversary has a high risk of losing its blocks when forking two honest blocks. 
Specifically, when two consecutive honest blocks appear, the adversary will adopt an honest block first. However, as $\alpha$ increases, the adversary will try to attack two blocks. Countermeasures are more effective on CHS when $\alpha$ is 0.3, the chain quality of CHS-C is 1.3 times that of CHS, while 2CHS-C only achieves 1.1 times of the original quality.

\begin{finding}
2CHS-C and CHS-C achieve better censorship resilience than the original protocols. The countermeasures are more efficient for CHS when it comes to censorship resilience.
\end{finding}

As for the censorship experimental results, we can see that countermeasures cannot completely resist attacks, but they can mitigate the impact to some extent. This is because the adversary only cares about excluding more honest blocks and does not care about their losses. Therefore, regardless of whether it affects their interests, the adversary will launch attacks. But countermeasures increase the probability of their attack failing. When $\alpha$ equals 0.3, the censorship of CHS-C is 1.5 times that of CHS, which is a better result than the 1.1 times for 2CHS-C.
This is because, in CHS, adversaries can attack up to two blocks at a time, while 2CHS only has one, and CHS has a worse censorship resilience than 2CHS. Countermeasures selecting branches by length make it much more difficult for the adversary to attack two blocks in CHS-C. As a result, the countermeasures are more effective for CHS.

\subsection{Countermeasures for Streamlet}
\label{subsec:streamlet-countermeasure}
The root cause of preemptive forking attacks in Streamlet is that the adversary can strategically publish its hidden certified blocks to prevent subsequent honest blocks from being certified. Due to network delays, it is difficult to ensure that all nodes (especially the leader) know the certified block from adversarial leaders before entering the next views. This is because the leader can strategically delay publishing its proposals to collect votes from a small partition of honest nodes by the end of the view. 
Therefore, without modifying the \texttt{voting rule}, \ie, a node only votes for the first block from the leader that extends the longest certified chain in its local view, we cannot effectively thwart attacks in Streamlet. 
However, modifying the \texttt{voting rule} is against our principles for countermeasures. 

\subsection{How to Design Chained BFT Protocols with Better Leadership Democracy?} The analysis results illustrate that these countermeasures can significantly improve the leadership democracy of the chained BFT protocols against various attacks. Here, we summarize some insights for designing chained BFT protocols with better leadership democracy.

\begin{packeditemize}
\item \textbf{Design components matter a lot.} 
The impact of design choices on leadership democracy is profound. Even a slight change can yield significant leadership democracy enhancements. For example, the countermeasure of FHS lies in the \textit{voting pattern}, a design component overlooked in the original design. By replacing DV with LBV, we can prevent the adversary from hiding QCs of honest blocks, making FHS achieve optimal censorship resilience. 
As another example, the countermeasures of 2CHS and CHS introduce uncertainty in the forking process, which can effectively reduce the potential losses of honest nodes. Interestingly, this change only slightly modifies the \texttt{proposing rule} (without affecting anything else), thereby having no impact on the security and performance of the protocol. 
The above insights also reveal that the design of the chained BFT protocol requires a thorough evaluation, which is what our framework provides.

\item \textbf{No one-size-fits-all chained BFT protocols.}
None of these protocols can achieve linear message complexity, responsiveness, and optimal censorship resilience due to the trade-offs between leadership democracy and system efficiency. 
2CHS-C and CHS-C enjoy linear complexity and low latency, but cannot achieve optimal censorship resilience. By contrast, FHS-C achieves optimal censorship resilience but suffers from additional overhead (in collecting the proof of the latest QC and additional communication steps).
\end{packeditemize}

\section{Related Work} \label{sec:related}
We review prior works on chained BFT protocols and MDP-based models for analyzing consensus protocols. To the best of our knowledge, we are the first to provide a systematic analysis of leadership democracy for chained BFT protocols by using MDP.

\bheading{Chained BFT protocols.}
Tendermint~\cite{Buchman2016TendermintBF} is among the first to realize frequent leader rotation for better leadership fairness. However, it does not use chaining, so we do not evaluate it in this work. 
Later, Buterin and Griffith propose Casper FFG~\cite{casper}, which combines leader rotation with chaining. 
The chaining structure can pipeline the multiple consensus phases of BFT consensus, improving efficiency and simplifying protocol design. 
A variant of Casper FFG is called Two-chain HotStuff (2CHS), formally described by Yin~\etal~\cite{hotstuffArxiv}. 2CHS can also be viewed as the chaining version of Tendermint. 
Despite the efficiency improvement, 2CHS (or Casper FFG) does not have responsiveness. 
To address this, chained HotStuff~\cite{hotstuffPODC} adds an extra phase of message exchange to achieve responsiveness and linear message complexity.
However, the additional phase increases the latency for commitment. 
Later, Fast-HotStuff (FHS)~\cite{fastHotStuff} includes proof of the latest QC in its block to achieve responsiveness, while introducing no additional delay compared with 2CHS. 
Unlike them for efficiency, Streamlet~\cite{chan2020streamlet} is proposed for simplifying consensus design, and so adopts different consensus rules.

Some variants of these pioneering protocols emerged afterward. Jolteon~\cite{gelashvili2022jolteon} and DiemBFT\_v4~\cite{baudet2019state} use quadratic view change and can commit blocks in two phases under steady state. 
Wendy~\cite{giridharan2021no} uses the no-commit proof for achieving optimal latency, optimistic responsiveness, and linear message complexity.
Marlin~\cite{sui2022marlin} uses rank to circumvent forking attacks and introduces virtual/shadow blocks to reduce latency and bandwidth. BeeGees~\cite{giridharan2023beegees} allow nodes to commit blocks without several consecutive blocks. 
In this work, we choose 2CHS, CHS, Streamlet, and FHS for evaluation because they not only inspire many subsequent chained BFT design~\cite{danezis2022narwhal, Kauri, gai2023scaling, sheng2021bft, stathakopoulou2022state, neu2021ebb, decouchant2022damysus}, but also are adopted in many blockchain platforms~\cite{ethereumcasper,kwon2016cosmos}.
More importantly, the framework can also be extended to support other chained BFT protocols due to their similarities in the chaining structure.

\bheading{Efficiency analysis and evaluation of chained BFT protocols.} 
Most existing analyses and experimental evaluations of chained BFT protocols focus on efficiency (\eg, throughput and latency), rather than leadership democracy. 
Several benchmark studies~\cite{dinh2017blockbench,shapiro2020performance,amiri2022bedrock,gramoli2023diablo, gai2021dissecting} evaluate the system efficiency of chained BFT protocols under various experimental settings.
Apart from these benchmarks, Niu et al.~\cite{niu2021performance, niu2021performance-tdsc} theoretically analyze the efficiency of CHS and propose two attacks: the delay attack and the forking attack. 
Cohen~\etal~\cite{cohen2022aware} propose an attack that can reduce the throughput of the chained HotStuff by over 30x and increase the latency by 5x in a setting of 1-3 Byzantine nodes out of 10 nodes.  
Giridharan~\etal~\cite{giridharan2023beegees} propose a liveness attack of CHS, \ie, making no block committed.  

Compared to these papers on evaluating efficiency, this paper provides a systematic analysis of the leadership democracy metrics, revealing insights into the incentive aspects of chained BFT protocols. In contrast to studies that focus on efficiency, we introduce the first MDP-based framework that derives optimal fairness attacks across the entire family of chained BFT protocols (2CHS, CHS, FHS, and Streamlet), uncovering previously unreported vulnerability patterns.

\bheading{MDP modeling of consensus.}
Despite being widely used for analyzing Nakamoto-style consensus protocols, MDP is never used for analyzing chained BFT protocols.
Previous works~\cite{sapirshtein2017optimal, gervais2016security, zhang2017necessity, bitcoin-ng} model selfish mining attacks and double spending attacks in PoW-based Nakamoto-style consensus protocols using MDP.
Zhang and Preneel~\cite{Zhang2019CommonMetrics} extend those works to a multi-metric framework for quantitatively analyzing PoW-based Nakamoto-style protocols against various attacks.
Adapting these studies to chained BFT protocols directly is difficult due to the fundamental difference between their designs, \eg, \texttt{voting} and \texttt{committing} rules. 

\section{Conclusion} \label{sec:conclusion}
In this paper, we develop a unified framework to analyze the leader democracy, including chain quality and censorship resilience, of chained BFT protocols. Using this framework, we evaluate four representative chained BFT protocols.
The evaluation results indicate that leader rotation is not enough to provide the leadership democracy guarantee; an adversary could utilize the existing design (\eg, chaining structure and voting pattern) to deteriorate the leadership democracy significantly. 
We further propose practical countermeasures with as few modifications as possible to resolve the found weakness of these protocols.
We conclude that even small changes in design components exert a profound impact on leadership democracy, which demonstrates the importance of fully evaluating protocols. Our work advocates the need for more attention on leadership democracy and the importance of systematic evaluation to explore potential attacks, understand the impact of design components, and make fair comparisons.

\normalem
\bibliographystyle{IEEEtran}
\bibliography{reference}

\vskip -2\baselineskip plus -1 fil
\begin{IEEEbiography}[{\includegraphics[width=1in,height=1.25in,clip,keepaspectratio]{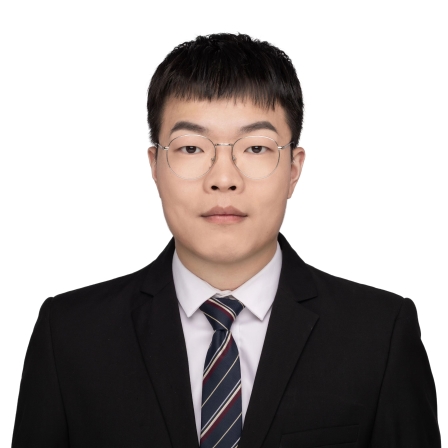}}] {Jianyu Niu} is currently a Research Fellow at the City University of Hong Kong. He received his Ph.D. degree from the University of British Columbia (Okanagan), Canada in 2021. He was a Research Associate and subsequently a Research Assistant Professor at the Southern University of Science and Technology from 2021 to 2025. His research interests include distributed systems, blockchain, and confidential computing.
\end{IEEEbiography}

\vskip -2\baselineskip plus -1 fil
\begin{IEEEbiography}[{\includegraphics[width=1in,height=1in,clip,keepaspectratio]{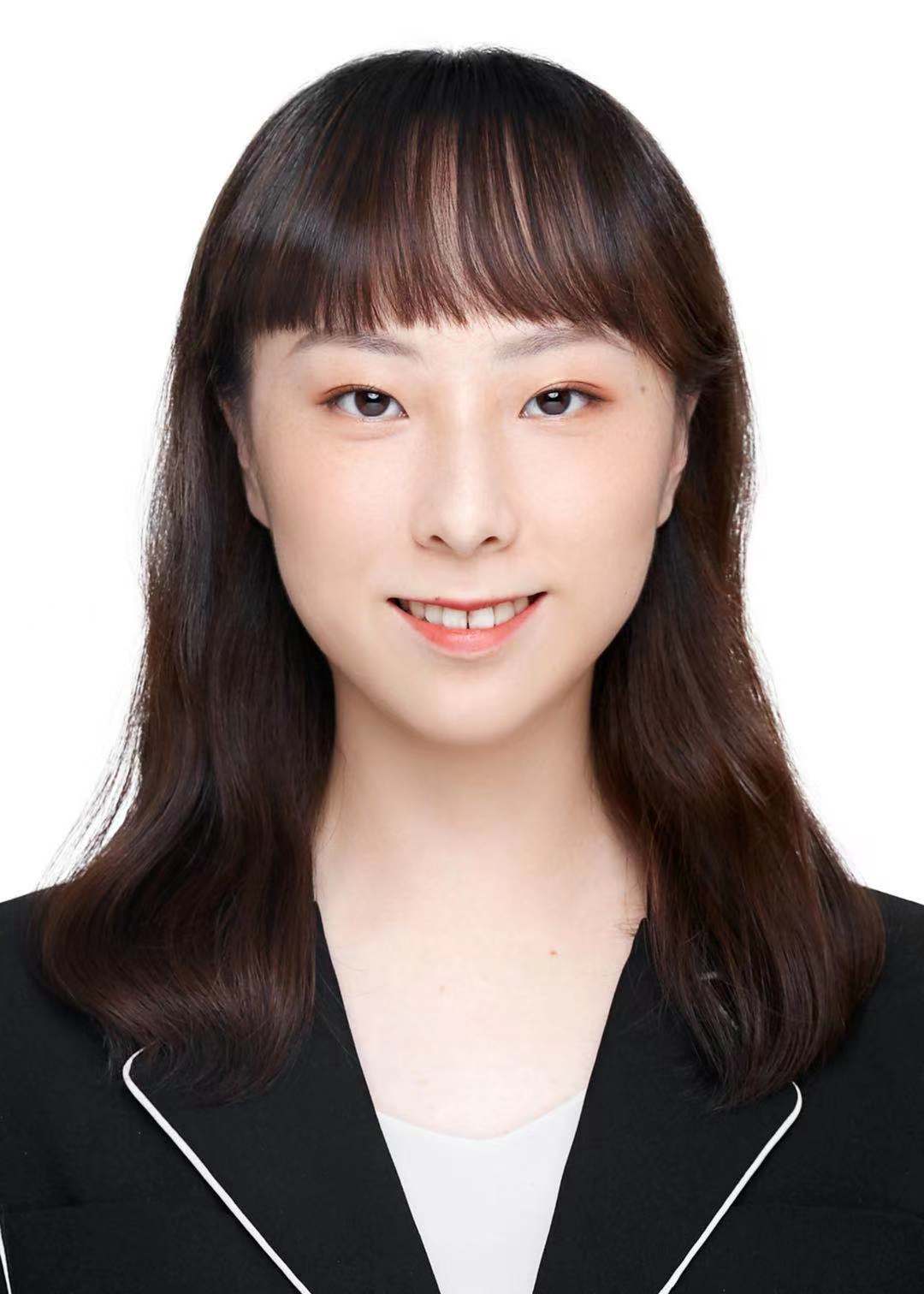}}] {Yining Tang} is currently pursuing the M.E. degree from Southern University of Science and Technology. She received her B.E. degree from the Southern University of Science and Technology. Her research interests include blockchain and trusted execution environments.
\end{IEEEbiography}

\vskip -2\baselineskip plus -1 fil
\begin{IEEEbiography}[{\includegraphics[width=1in,height=1.15in,clip,keepaspectratio]{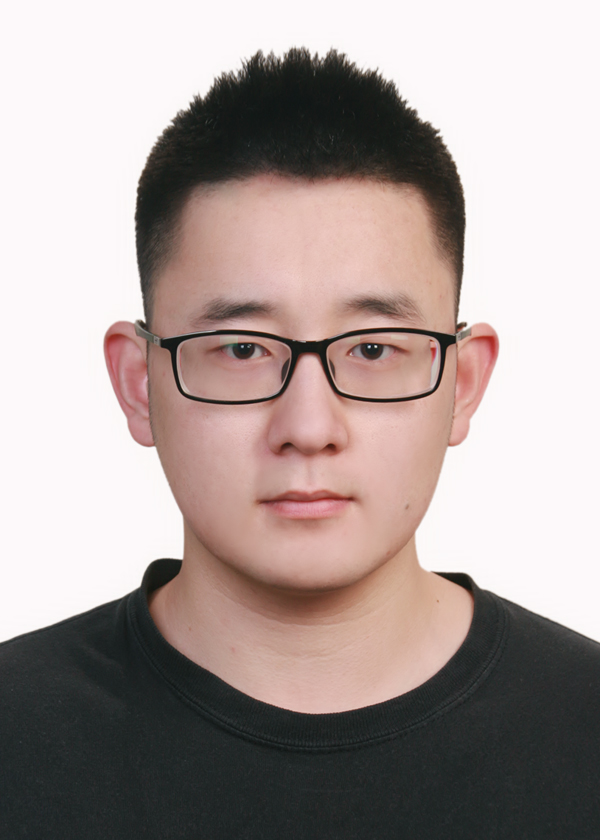}}]{Runchao Han} is currently a Researcher and Engineer at Babylon Labs. He completes his PhD at Monash University and CSIRO's Data61, Australia. He received an MSc degree from The University of Manchester, United Kingdom, and a bachelor degree from Beijing University of Posts and Telecommunications, China.
His research focuses on distributed systems, especially security and scalability issues in blockchains. 
\end{IEEEbiography}

\vskip -2\baselineskip plus -1 fil
\begin{IEEEbiography}[{\includegraphics[width=1in,height=1.25in,clip,keepaspectratio]{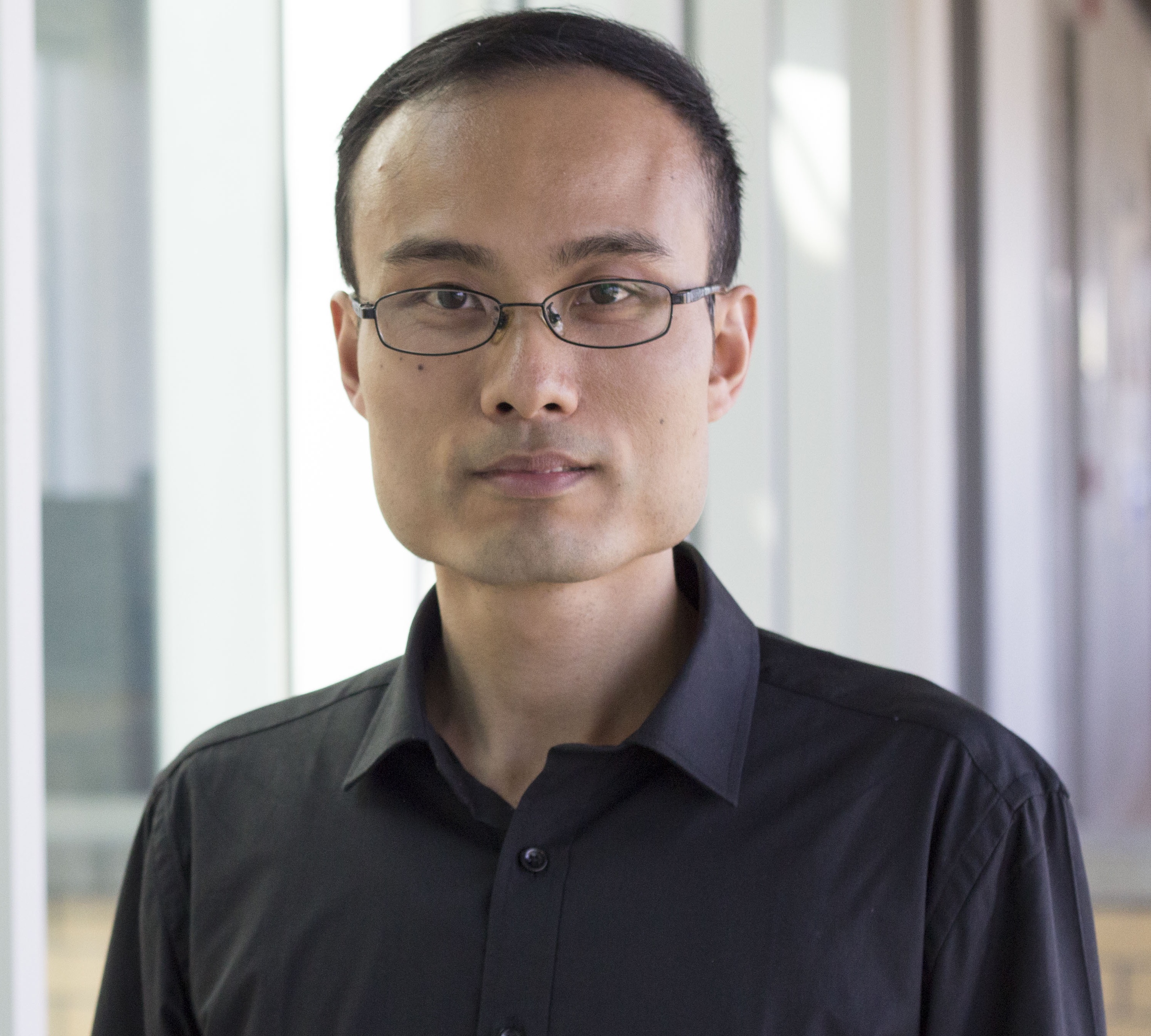}}]{Chen Feng}
    received the Ph.D. degrees from The University of Toronto, Canada, in 2009 and 2014, respectively. From 2014 to 2015, he was a Postdoctoral Fellow with Boston University, USA, and EPFL, Switzerland.

    He joined the School of Engineering, University of British Columbia (Okanagan Campus), Kelowna, Canada, in July 2015, where he is currently an Associate Professor. He is a co-cluster lead of Blockchain@UBC and Principal’s Research Chair in Blockchain. He is interested in adapting new ideas and tools from information theory, coding theory, stochastic processes, and optimization to design better communication networks, with a particular emphasis on blockchain technology.
\end{IEEEbiography}

\vskip -2\baselineskip plus -1 fil
\begin{IEEEbiography}[{\includegraphics[width=1in,height=1.25in,clip,keepaspectratio]{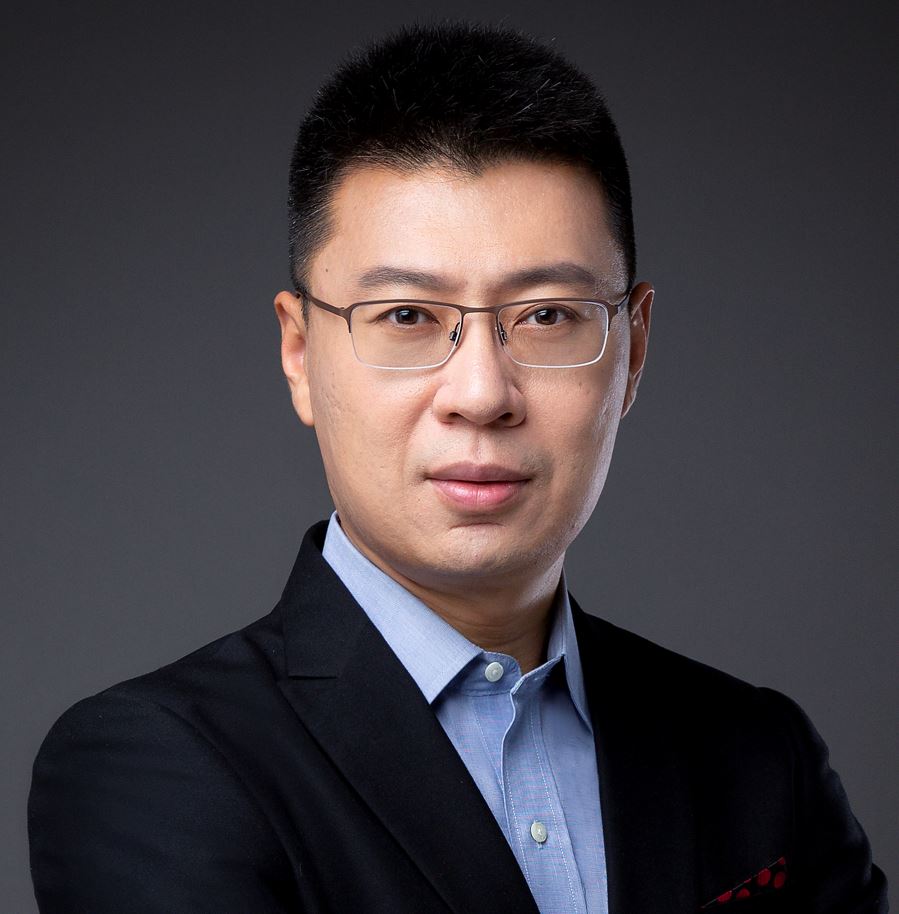}}]{Yinqian Zhang} is a Professor in the Department of Computer Science and Engineering at Southern University of Science and Technology. His research interests lie in system security, including side channels, trusted and confidential computing, and cloud security. 
\end{IEEEbiography}
\clearpage
\appendix
\subsection{Discussion}\label{appen:discussion}
\subsubsection{Leadership Democracy during Asynchrony} \label{appen:asynchrnoy}
This paper focuses on analyzing leadership democracy when the network is synchronous (\ie, after GST). The results already show the degradation of leadership democracy under attack and also reveal some insights for building better chained BFT protocols. We also observe that when the network is asynchronous, the adversary can further deteriorate the leadership democracy. This is because during asynchrony, an honest block and its associated QC may not be received by all honest nodes at the end of their generation view. 
For example, if an honest block that extends the longest certified chain is proposed during synchrony in Streamlet, it will be certified and received by all honest nodes. 
Then, all subsequent blocks have to extend it. Otherwise, they cannot obtain votes from honest nodes due to the \texttt{voting rule} (\appref{appen:Streamlet}).  
However, during asynchrony, the honest block may not be certified or seen by all honest nodes, so the adversary still has the chance to propose a block to override it.  

\subsubsection{Leadership Democracy under Round Robin Policy} \label{appen:robin}
This paper focuses on the random leader election policy since it is commonly used in blockchains, especially permissionless ones~\cite{schwarz2022three, aptos, cypherium, Hotshot}. Besides, the round-robin policy is another popular one, where the leader in each view is pre-determined rather than randomly selected. 
This public leadership assignment allows the adversary to deteriorate the leadership democracy.   
For example, in 2CHS, an adversary can create forks to override one honest block. 
Thus, the optimal strategy for the adversary is to corrupt the leader every other, such that every adversarial block can override one honest block.
Existing analysis has shown similar conclusions~\cite{fastHotStuff, giridharan2023beegees}, where the adversary can violate the liveness of CHS using the round-robin policy, while only increasing the latency of CHS using the random policy~\cite{gai2021dissecting, niu2021performance}. 

\subsubsection{Future Directions for Chained BFT Consensus} \label{appen:futureWork}
We provide several future directions for protocol design.

\bheading{Evaluating design components under attack.}
Our work advocates the need for more attention to leadership democracy. In particular, our evaluation results illustrate that existing chained BFT protocols have not fully considered the impact of design components on leadership democracy, especially under attack. 
For example, CHS adds a phase to achieve responsiveness, but sacrifices the chain quality and censorship resilience under attack.
Our work shows that such a cost is not fundamental -- the random \texttt{proposing rule} proposed in \S\ref{sec:countermeasure} can improve these two metrics without reducing any efficiency of CHS. A systematic evaluation during design can guide researchers to develop chained BFT protocols with better leadership democracy.

\bheading{Modularizing BFT protocol design.}
Unlike existing efforts that propose groundbreaking designs for chained BFT protocols, our work has explored a modular approach so that one can develop a new design by simply composing existing off-the-shelf components. Indeed, all the countermeasures proposed in \S\ref{sec:countermeasure} remain modular because they only replace an off-the-shelf component with another one.
Due to the modular design, such changes do not reduce any security or efficiency of chained BFT protocols, while at the same time effectively enhancing the chain quality and censorship resilience.
This highlights the power of modular design, sharing similar principles with the efforts on adding ``gadgets" on top of blockchains~\cite{neu2022availability, neu2021ebb, casper}.

\bheading{Making forking attacks accountable.}
Apart from the countermeasures, an orthogonal approach to improve leadership democracy is to make forking attacks on chain quality and censorship resilience accountable.
That is, upon a forking attack, the honest nodes in the protocol can irrefutably identify a set of adversarial nodes who are launching the attack.
Consequently, the system can penalize the identified adversarial nodes, such as removing them from the system, applying financial penalties, and seeking external regulations.
Existing literature focuses on the accountability of safety attacks~\cite{haeberlen2007peerreview,wang2021accountability,sheng2021bft,neu2022availability} with little attention to chain quality or censorship resilience attacks.

However, detecting chain quality or censorship resilience attacks can still be challenging in many cases.
For example, it is possible for an honest node to be absent for some views and to vote for a fork, due to the network delay.
However, these behaviors are the preliminary steps to launch forking attacks on the chain quality and censorship resilience.
The indistinguishability between adversarial behaviors and random network delay make it challenging to identify forking attacks.
Besides, it is more challenging when deploying chained BFT under permissionless blockchains where nodes' identities are anonymous.

\subsection{Detailed Protocol Description} \label{appen:chainedBFT}
There are similarities among these chained BFT protocols, such as the basic principle similar to BFT consensus. Moreover, they adopt the chain structure to improve throughput and simplify the protocol design. We select four classic chained BFT protocols, each with its own characteristics. Many subsequent protocols are based on these protocols.

\subsubsection{Two-chain HotStuff} \label{appen:2chs}
\begin{figure}[t]
    \centering
    \begin{subfigure}[b]{0.45\linewidth}
        \centering
        \includegraphics[width=\linewidth]{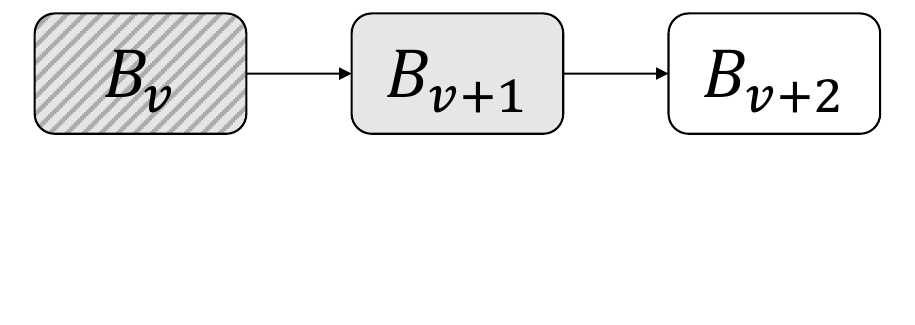}
        \caption{The protocol rules.}
        \label{fig:2CHS-rule}
    \end{subfigure} 
    \begin{subfigure}[b]{0.45\linewidth}
        \centering   
        \includegraphics[width=0.95\linewidth]{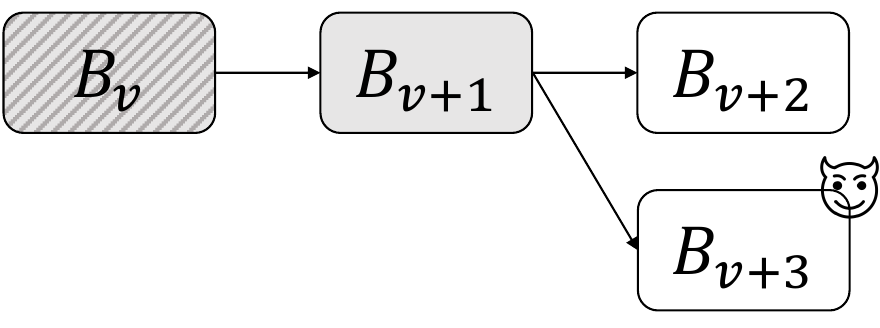}
        \caption{The attack scenario.}
        \label{fig:2CHS-attack}
    \end{subfigure}
    \label{fig:2CHS-detail}
    \caption{Detailed protocol description of 2CHS and its attack scenario.}
    \vspace{-5mm}
\end{figure}

Two-chain HotStuff originates from Casper FFG, which adds chaining based on Tendermint and extends to a two-chain locking mechanism. The following are the rules of 2CHS.
\begin{packeditemize}
    \item \texttt{Proposing rule.} The leader proposes a block $B$ to extend the highest certified block.

    \item \texttt{Voting rule.} There are two noteworthy views: $i$) $last\_voted\_view$, the last view for which the node voted, and $ii$) $locked\_view$, the highest known view number of the parent block. For example, in \figref{fig:2CHS-rule}, when a node first receives block $B_{v+2}$ in view $v+2$ and votes for it, its $last\_voted\_view$ is updated to $v+2$. Since it makes all nodes see $B_{v+1}.QC$, the highest parent block is $B_{v+1}$, and so $locked\_view$ is view $v+1$. 
    A block satisfies the \texttt{voting rule} only if: $i$) its view number is greater than $last\_voted\_view$, and $ii$) the view number of its parent block is at least $locked\_view$. 
    
    \item \texttt{Committing rule.} If there are two blocks $B_v$ and $B_{v+1}$ proposed in two consecutive views $v$ and $v+1$, and an additional block extends the block $B_{v+1}$, nodes will commit the block $B_v$ and all its preceding blocks. The blocks $B_v$ and $B_{v+1}$ are referred to as $2$-direct chain.
\end{packeditemize}

\bheading{Forking attack in 2CHS.} The attack scenario in 2CHS is shown in \figref{fig:2CHS-attack}, where $B_{v+3}$ extends $B_{v+1}$ and is proposed at the same height as $B_{v+2}$. Since the $locked\_view$ at this time is $B_{v+1}$, $B_{v+3}$ meets the \texttt{voting rule} of other nodes: $i$) its view number $v+3$ is greater than $last\_voted\_view$ $v+2$; $ii$) the view number of its parent block equals the locked view number. Therefore, $B_{v+3}$ will receive enough votes, and the next leader will extend the highest certified block, and thus $B_{v+3}$ overrides $B_{v+2}$.

\subsubsection{Chained HotStuff} \label{appen:3chsHotStuff} 
Chained HotStuff (CHS)~\cite{hotstuffPODC} creatively adopts a three-chain \texttt{committing rule}  (rather than the two-phase \texttt{committing rule}) to enable the protocol to reach consensus at the pace of actual network delay. 
The significant difference between the rules of 2CHS and CHS is the $locked\_view$. More precisely, $locked\_view$ is the highest known view number of the grandparent block in CHS, while is the highest known view number of the parent block in 2CHS.

\begin{figure}[t]
    \centering
    \includegraphics[width=0.57\linewidth]{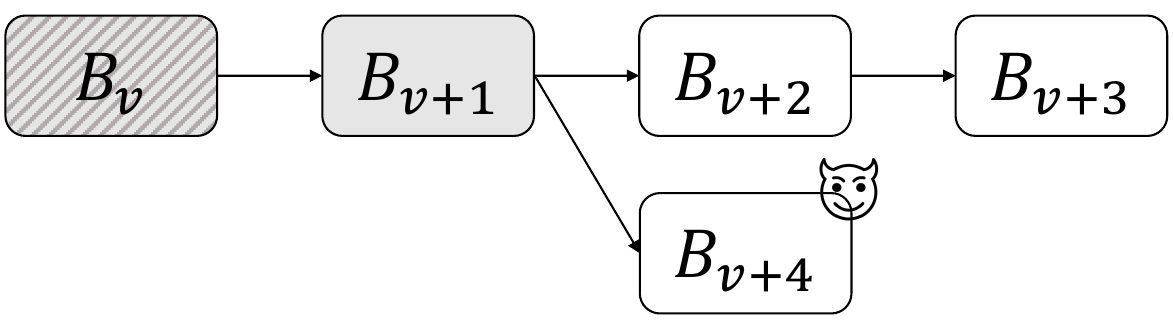}
    \caption{The chained blocks in CHS. Block $B_v$, $B_{v+1}$ and $B_{v+2}$ are consecutive blocks. Nodes will commit $B_v$ when receiving proposal of $B_{v+3}$, and lock at $B_{v+1}$. The adversarial block $B_{v+4}$ could override two honest blocks.}
    \label{fig:CHS-example}
\end{figure}

\begin{packeditemize}
\item \texttt{Proposing rule.} Leader proposes a block $B$ to extend the highest certified block.

\item \texttt{Voting rule.} Nodes still maintain two parameters of the view number: The $last\_voted\_view$ remains the same as 2CHS, while $locked\_view$ becomes the highest known view number of the grandparent block. For example, in \figref{fig:CHS-example}, when a node first receives the block $B_{v+3}$ and votes for it, its $last\_voted\_view$ is updated to $v+3$, and the highest grandparent block becomes $B_{v+1}$, and so $locked\_view$ is $v+1$. 
A block satisfies the \texttt{voting rule} only if: $i$) its view number is greater than $last\_voted\_view$, and ($ii$) the view number of its grandparent block is at least $locked\_view$. 

\item \texttt{Committing rule.} If there are three blocks $B_v$, $B_{v+1}$ and $B_{v+2}$ proposed in three consecutive views $v$, $v+1$, and $v+2$, and an additional block extends block $B_{v+2}$ ($B_{v+3}$ in \figref{fig:CHS-example}), nodes will commit block $B_v$ and all its preceding blocks. The first three consecutive blocks are referred to as the $3$-direct chain.
\end{packeditemize}

\bheading{Forking attack in CHS.} The attack scenario of CHS is shown in \figref{fig:CHS-example}, where the leader of the view $v+4$ is an adversary and proposes a block extending $B_{v+1}$.
$B_{v+4}$ meets the two conditions of the \texttt{voting rule}. Therefore, other honest nodes will vote for it, and the subsequent leader will propose a block after it based on the \texttt{proposing rule}. In this way, $B_{v+4}$ will eventually replace $B_{v+2}$ and $B_{v+3}$ on the chain.

\subsubsection{Fast-HotStuff} \label{appen:fasthotstuff}
Fast-HotStuff~\cite{fastHotStuff} has lower latency compared to 2CHS and is resilient to a forking attack. Fast-HotStuff adds a small overhead to the block during an unhappy path (when the primary fails). The protocol rules are as follows:
\begin{packeditemize}
\item \texttt{Proposing rule.} A leader proposes a block $B$ to extend the highest certified block according to the included $(n-f)$ QC (proof of the latest QC).

\item \texttt{Voting rule.} If a leader extends the highest certified block by checking the proof of the latest QC, nodes vote for it.

\item \texttt{Committing rule.} If there are two blocks $B_v$ and $B_{v+1}$ proposed in two consecutive views $v$, $v+1$, and an additional block extends $B_{v+1}$ (broadcasting $B_{v+1}.QC$ to other nodes), nodes will commit the first block $B_v$. 
\end{packeditemize}

\bheading{Forking attacks in FHS.} 
The attack modes of both FHS and 2CHS belong to posterior forking attacks and result in an adversary leader forking an honest block. But the reasons for their attacks are not completely the same. 
FHS claims to be robust against forking attacks because proof of the latest QC is required when proposing blocks. A leader has to provide the aggregation of the latest QC seen by other nodes when proposing a block, and the proposed block must extend the corresponding or higher block of the latest QC. 

\begin{figure}[t]
    \centering
    \includegraphics[width=0.45\linewidth]{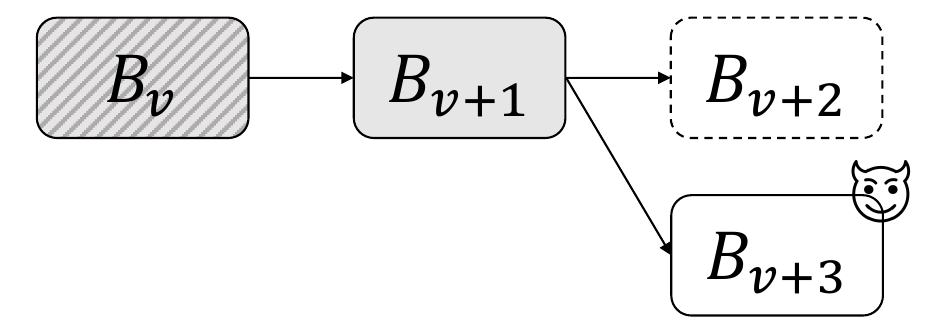}
    \caption{The chained blocks in FHS. Block $B_v$ and $B_{v+1}$ are consecutive blocks. Nodes will commit $B_v$ when receiving proposal of $B_{v+2}$, and lock at $B_{v+1}$. The adversarial block $B_{v+3}$ could override one honest block.}
    \label{fig:FHS-attack}
\end{figure}

With this framework, we discover some new adversarial behavior. The \texttt{voting rule} of FHS is to send votes to the next leader, so when the next leader collects the latest QC of other nodes, it can pretend not to have collected enough votes to form a QC and hinder the formation of an honest block. As shown in \figref{fig:FHS-attack}, the honest leader of view $v+2$ proposes $B_{v+2}$ after $B_{v+1}$, it provides the proof of latest QC $B_{v+1}.QC$. Other nodes vote for it and send the vote to the adversary leader of $v+3$. The adversary leader proposes a block after $B_{v+1}$ instead of publishing $B_{v+2}.QC$, indirectly forking $B_{v+2}$.

\subsubsection{Streamlet} \label{appen:Streamlet}
Streamlet~\cite{chan2020streamlet}, Chan \etal proposed a simple block proposing and \texttt{voting rule}: the longest certified chain rule. Streamlet is built based on Streamlined blockchains~\cite{shi2019streamlined}, and has simplified the protocols by removing the notion of freshness. 

\begin{packeditemize}
    \item \texttt{Proposing rule}. The leader proposes a block built on top of the longest certified chain.~\footnote{In Streamlet~\cite{shi2019streamlined}, a certified block is called notarized block. Here, we use a certified block to be consistent with the description in other protocols}

    \item \texttt{Voting rule}. A node will vote for the first proposal if the proposed block is built on top of the longest \textit{certified} chain it has seen. Note that the vote is broadcast.

    \item \texttt{Committing rule}. Whenever three blocks proposed in three consecutive views get certified, the first two blocks out of the three along with the ancestor blocks are committed.
\end{packeditemize}

\begin{figure}[t]
    \centering
    \includegraphics[width=0.45\linewidth]{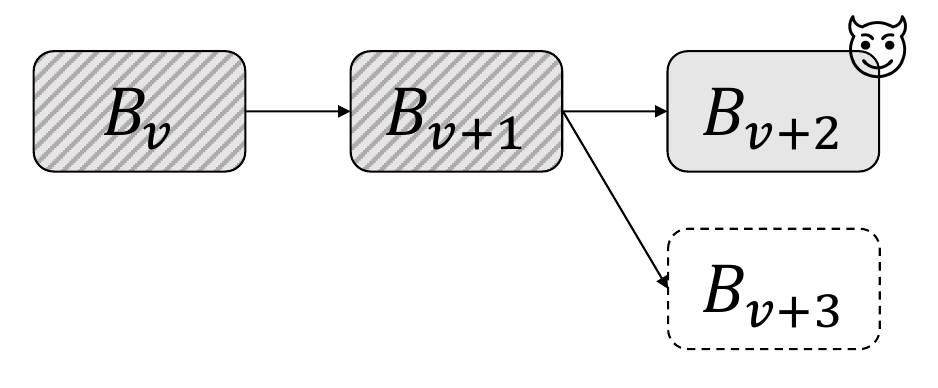}
    \caption{The chained blocks in Streamlet. Block $B_v$, $B_{v+1}$ and $B_{v+2}$ are consecutive blocks. Nodes will finalize $B_v$ and $B_{v+1}$ when receiving QC (also called notarization in Streamlet) of $B_{v+2}$. The adversarial block $B_{v+2}$ could invalidate the honest block $B_{v+3}$.}
    \label{fig:Streamlet-attack}
\end{figure}

\bheading{Forking attacks in Streamlet.} Different from other protocols, Streamlet follows the longest certified chain rule. Therefore, the adversary cannot override the generated honest block. However, the adversary can strategically delay its block proposal, make a subset of at least $f+1$ but less than $2f+1$ honest nodes see the block, and control the voting of all adversarial nodes. This could result in a delay in certifying the block until the next honest leader suggests a new block.
As shown in \figref{fig:Streamlet-attack}, the adversaries temporarily withhold their votes for the adversarial block $B_{v+2}$, and $B_{v+2}$ could not get enough votes. The leader of view $v+3$ is not aware of $B_{v+2}$ and will propose a block that extends $B_{v+1}$. After that, all nodes receive votes from adversaries for $B_{v+2}$, causing $B_{v+2}$ to be certified. This results in a block being certified at the same height as $B_{v+3}$. Once other nodes receive $B_{v+3}$, they will not vote for it.

\subsection{MDP model of 2CHS, CHS, FHS and Streamlet} \label{appen:mdpmodel}

\bheading{CHS}
The state transition and reward matrices for CHS are shown in \tabref{tab:state_trans_CHS}. The only difference from 2CHS is that the maximum value of \lh is $2$.

\begin{table}[t]
\caption{State transition and reward matrices for CHS.} 
\label{tab:state_trans_CHS}
\scriptsize
\centering
\ra{1.1}
\resizebox{\linewidth}{!}{%
\begin{tabular}{@{}cccc@{}}
\toprule[1pt]
\textbf{State $\times$ Action} & \textbf{Resulting State} & \textbf{Probability} & \textbf{Reward} \\ \midrule[1pt]
\multirow{2}{*}{$(\la, \lh, 1)$, \texttt{Adopt}} & $(0, 1, 0)$ & $\alpha$ & \multirow{2}{*}{$(\lh, 0, 0)$}\\
 & $(0, 1, 1)$ & $1-\alpha$ &  \\ \midrule[1pt]
\multirow{2}{*}{$(\la, \lh, 0)$, \texttt{Adopt}} & $(1, 0, 0)$ & $\alpha$ & \multirow{2}{*}{$(\lh, 0, 0)$} \\
 & $(1, 0, 1)$ & $1-\alpha$ & \\ \midrule[1pt]
\multirow{2}{*}{$(\la, \lh, 1)$, \texttt{Wait}} & $(0, \min(\lh+1,\textbf{2}), 0)$ & $\alpha$ & \multirow{2}{*}{\shortstack{$(1, 0, 0)$ if $\textbf{\lh=2}$\\$(0, 0, 0)$ otherwise}} \\
 & $(0, \min(\lh+1,\textbf{2}), 1)$ & $1-\alpha$ & \\ \midrule[1pt]
\multirow{2}{*}{$(0, \lh, 0)$, \texttt{Wait}} & $(1, \lh, 0)$ & $\alpha$ & \multirow{2}{*}{$(0, 0, 0)$} \\
 & $(1, \lh, 1)$ & $1-\alpha$ & \\ \midrule[1pt]
\multirow{2}{*}{$(1, \lh, 0)$, \texttt{Wait}} & $(1, 0, 0)$ & $\alpha$ & \multirow{2}{*}{$(0, 1, \lh)$} \\
 & $(1, 0, 1)$ & $1-\alpha$ & \\ \midrule[1pt]
\multirow{2}{*}{$(1, \lh, 1)$, \texttt{Release}} & $(0, 1, 0)$ & $\alpha$ & \multirow{2}{*}{$(0, 1, \lh)$} \\
 & $(0, 1, 1)$ & $1-\alpha$ & \\ \midrule[1pt]
\multirow{2}{*}{$(1, \lh, 0)$, \texttt{Release}} & $(1, 0, 0)$ & $\alpha$ & \multirow{2}{*}{$(0, 1, \lh)$} \\
 & $(1, 0, 1)$ & $1-\alpha$ & \\ \midrule[1pt]
\end{tabular}%
}
\end{table}

\bheading{MDP model of Streamlet}
The state transition and reward matrices for Streamlet are shown in \tabref{tab:state_trans_Streamlet}. 
There is an additional action called \texttt{withhold}, which means that after the current leader proposes a block, the hidden adversarial block will be revealed. This action can only be taken when there is a hidden adversarial block.


\begin{table}[t]
\caption{State transition and reward matrices for Streamlet. 
The variable $\alpha$ denotes the fraction of adversarial nodes. The \texttt{release} and \texttt{withhold} actions are feasible only when $\la>0$. The reward is a tuple of $(B_h, B_a, O_h)$.} 
\label{tab:state_trans_Streamlet}
\tiny
\centering
\resizebox{\linewidth}{!}{%
\begin{tabular}{@{}cccc@{}}
\toprule[1pt]
\textbf{State $\times$ Action} & \textbf{Resulting State} & \textbf{Probability} & \textbf{Reward} \\ \midrule[1pt]
\multirow{2}{*}{$(\la, \lh, 1)$, \texttt{Adopt}} & $(0, 1, 0)$ & $\alpha$ & \multirow{2}{*}{$(\lh, 0, 0)$}\\
 & $(0, 1, 1)$ & $1-\alpha$ &  \\ \midrule[1pt]
\multirow{2}{*}{$(\la, \lh, 0)$, \texttt{Adopt}} & $(1, 0, 0)$ & $\alpha$ & \multirow{2}{*}{$(\lh, 0, 0)$} \\
 & $(1, 0, 1)$ & $1-\alpha$ & \\ \midrule[1pt]
\multirow{2}{*}{$(\la, \lh, 1)$, \texttt{Wait}} & $(0, \lh+1, 0)$ & $\alpha$ & \multirow{2}{*}{$(0, 0, 0)$} \\
 & $(0, \lh+1, 1)$ & $1-\alpha$ & \\ \midrule[1pt]
\multirow{2}{*}{\shortstack{$(0, \lh, 0)$, \texttt{Wait}\\ \lh$\!=$ 0}} & $(1, \lh, 0)$ & $\alpha$ & \multirow{2}{*}{$(0, 0, 0)$} \\
 & $(1, \lh, 1)$ & $1-\alpha$ & \\ \midrule[1pt]
\multirow{2}{*}{\shortstack{$(0, \lh, 0)$, \texttt{Wait}\\ \lh$\!>$ 0}} & $(0, \lh, 0)$ & $\alpha$ & \multirow{2}{*}{$(0, 0, 0)$} \\
 & $(0, \lh, 1)$ & $1-\alpha$ & \\ \midrule[1pt]
\multirow{2}{*}{$(1, \lh, 0)$, \texttt{Wait}} & $(1, 0, 0)$ & $\alpha$ & \multirow{2}{*}{$(\lh, 1, 0)$} \\
 & $(1, 0, 1)$ & $1-\alpha$ & \\ \midrule[1pt]
\multirow{2}{*}{$(1, \lh, 1)$, \texttt{Release}} & $(0, 1, 0)$ & $\alpha$ & \multirow{2}{*}{$(\lh, 1, 0)$} \\
 & $(0, 1, 1)$ & $1-\alpha$ & \\ \midrule[1pt]
\multirow{2}{*}{$(1, \lh, 0)$, \texttt{Release}} & $(1, 0, 0)$ & $\alpha$ & \multirow{2}{*}{$(\lh, 1, 0)$} \\
 & $(1, 0, 1)$ & $1-\alpha$ & \\ \midrule[1pt]
\multirow{2}{*}{$(1, \lh, 1)$, \texttt{Withhold}} & $(0, 0, 0)$ & $\alpha$ & \multirow{2}{*}{$(\lh, 1, 1)$} \\
 & $(0, 0, 1)$ & $1-\alpha$ & \\ \midrule[1pt]
\multirow{2}{*}{$(1, \lh, 0)$, \texttt{Withhold}} & $(1, 0, 0)$ & $\alpha$ & \multirow{2}{*}{$(\lh, 1, 0)$} \\
 & $(1, 0, 1)$ & $1-\alpha$ & \\ \midrule[1pt]
\end{tabular}%
}
\end{table}

\subsection{MDP model of 2CHS-C of and CHS-C} \label{appen:randomness}

\begin{table*}[t]
\caption{State transition and reward matrices for CHS-C.} 
\label{tab:state_trans_CHS-C}
\scriptsize
\centering
\begin{threeparttable}
\ra{1.1}
\begin{tabular}{@{}cccc@{}}
\toprule[1pt]
\textbf{State $\times$ Action} & \textbf{Resulting State} & \textbf{Probability} & \textbf{Reward} \\ \midrule[1pt]

\multirow{2}{*}{$(\la, \lh, \cnt, 1)$, \texttt{Adopt}} & $(0, 1, 1, 0)$ & $\alpha$ & \multirow{2}{*}{$(\lh, 0, 0)$}\\
 & $(0, 1, 1, 1)$ & $1-\alpha$ &  \\ \midrule[1pt]
 
\multirow{2}{*}{$(\la, \lh, \cnt, 0)$, \texttt{Adopt}} & $(1, 0, 0, 0)$ & $\alpha$ & \multirow{2}{*}{$(\lh, 0, 0)$} \\
 & $(1, 0, 0, 1)$ & $1-\alpha$ & \\ \midrule[1pt]
 
\multirow{2}{*}{\shortstack{$(\la, \lh, \cnt, 1)$, \texttt{Wait} \\ $\lh=0 \lor (\lh>0 \land \la\leq\lh)$}} & $(\la, \lh+1, \cnt+1,  0)$ & $\alpha$ & \multirow{2}{*}{$(0, 0, 0)$} \\
 & $(\la, \lh+1, \cnt+1, 1)$ & $1-\alpha$ & \\ \midrule[1pt]

\multirow{4}{*}{\shortstack{$(\la, \lh, \cnt, 1)$, \texttt{Wait} \\ $\la>\lh \land \lh>0$}} & $(\la, \lh+1, \cnt+1, 0)$ & $\gamma\alpha$ & \multirow{2}{*}{$(0, 0, 0)$} \\
 & $(\la, \lh+1, \cnt+1, 1)$ & $\gamma(1-\alpha)$ & \\
 & $(1, 1, 1, 0)$ & $(1-\gamma)\alpha$ & \multirow{2}{*}{$(0, \la-1, \lh)$} \\
 & $(1, 1, 1, 1)$ & $(1-\gamma)(1-\alpha)$ & \\ \midrule[1pt]

\multirow{2}{*}{\shortstack{$(\la, \lh, \cnt, 0)$, \texttt{Wait, Release} \\ $\lh=0 \lor (\la>\lh \land \lh>0)$}} & $(1, 0, 0, 0)$ & $\alpha$ & \multirow{2}{*}{$(0, \la, \lh)$} \\
 & $(1, 0, 0, 1)$ & $1-\alpha$ & \\ \midrule[1pt]
 
\multirow{2}{*}{\shortstack{$(\la, \lh, \cnt, 0)$, \texttt{Wait, Release} \\ $\la\leq\lh \land \lh>0$}} & $(\la+1, \lh, 0, 0)$ & $\alpha$ & \multirow{2}{*}{$(0, 0, 0)$} \\
 & $(\la+1, \lh, 0, 1)$ & $1-\alpha$ & \\ \midrule[1pt]

\multirow{2}{*}{\shortstack{$(\la, \lh, \cnt, 1)$, \texttt{Release} \\ $\lh=0 \lor (\la>\lh \land \lh>0)$}} & $(0, 1, 1, 0)$ & $\alpha$ & \multirow{2}{*}{$(0, \la, \lh)$} \\
& $(0, 1, 1, 1)$ & $1-\alpha$ & \\ \midrule[1pt]

\multirow{4}{*}{\shortstack{$(\la, \lh, \cnt, 1)$, \texttt{Release} \\ $\lh>0 \land \la=\lh$}} & $(\la, \lh+1, \cnt+1, 0)$ & $\gamma\alpha$ & \multirow{2}{*}{$(0, 0, 0)$} \\
 & $(\la, \lh+1, \cnt+1, 1)$ & $\gamma(1-\alpha)$ & \\
 & $(0, 1, 1, 0)$ & $(1-\gamma)\alpha$ & \multirow{2}{*}{$(0, \la, \lh)$} \\
 & $(0, 1, 1, 1)$ & $(1-\gamma)(1-\alpha)$ & \\ \midrule[1pt]

\multirow{2}{*}{\shortstack{$(\la, \lh, \cnt, 1)$, \texttt{Release} \\ $\lh>0 \land \la<\lh$}} & $(\la, \lh+1, \cnt+1, 0)$ & $\alpha$ & \multirow{2}{*}{$(0, 0, 0)$} \\
& $(\la, \lh+1, \cnt+1, 1)$ & $1-\alpha$ & \\ \midrule[1pt]
\end{tabular}
\begin{tablenotes}
    \item When $\cnt=2$, the resulting state $\cnt+1$ turns to be $\lh=2$, $\la=0$, $\cnt=2$, and the corresponding reward $B_h=\lh-1$.

\end{tablenotes}
\end{threeparttable}
\end{table*}

The state transition and reward matrices of CHS-C is shown in 
\tabref{tab:state_trans_CHS-C}. The matrices for 2CHS-C is the same as CHS-C except that the maximum value of \cnt is 1. The randomness in the \texttt{proposing rule} introduces unpredictability, by which the adversary may lose competition in some forks. 

{
\subsection{Occasional Honest Leader Timeouts} \label{append:gamma-expe}

To enhance the completeness of the model, we add an adjustable parameter $\gamma$ to the original MDP model. This addition aims to reflect the probability of block proposal failures by honest leaders. To better visualize the results, we set the Byzantine node fraction $\alpha$ to 0.3 and illustrate the variations of leadership democracy metrics across varying gamma values, namely 0, 10\%, and 20\%.

\begin{figure}[t]
    \centering
    \includegraphics[width=\linewidth]{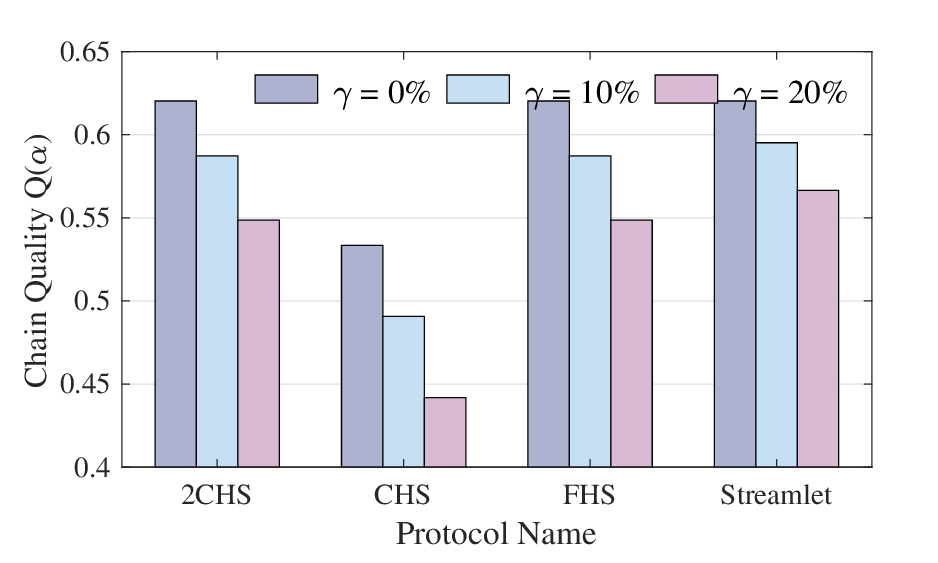}
    \caption{Chain quality of four protocols under different $\gamma$ when $\alpha=0.3$.}
    \label{fig:honest_fail_Quality}
\end{figure}

\begin{figure}[t]
    \centering
    \includegraphics[width=\linewidth]{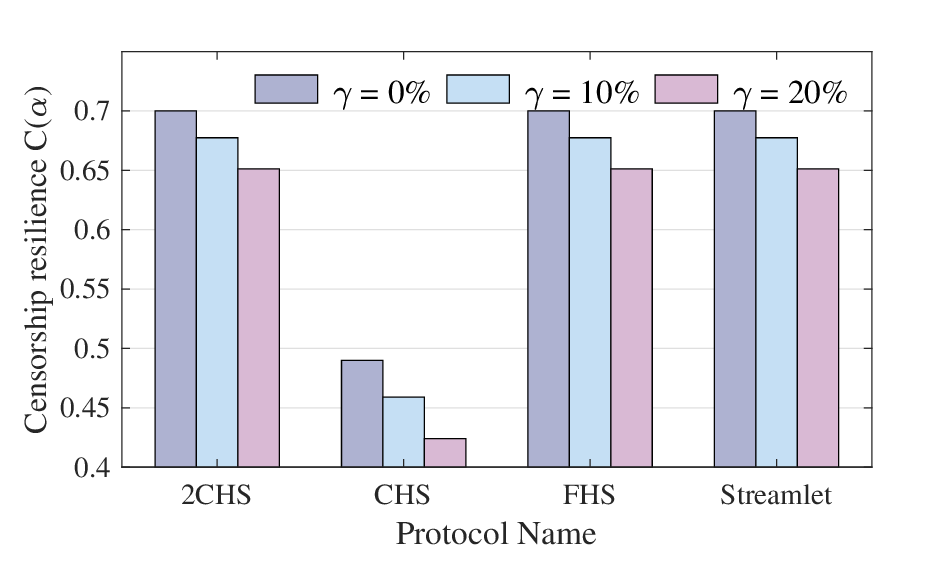}
    \caption{Censorship resilience of four protocols under different $\gamma$ when $\alpha=0.3$.}
    \label{fig:honest_fail_Censor}
\end{figure}

The evaluation results are shown in \figref{fig:honest_fail_Quality} and \figref{fig:honest_fail_Censor}. In our experiments with four protocols, we observe that as $\gamma$ increases, both chain quality and censorship resilience decline across all protocols. This indicates that occasional honest leader timeouts can further weaken the protocol's leadership democracy. The decline in censorship resilience is less pronounced compared to the decline in chain quality for most protocols, indicating that censorship resilience is more robust to honest leader timeouts. The use of different $\gamma$ values allows for a sensitivity analysis of the protocols under varying degrees of honest leader timeouts. This provides valuable information to understand how each protocol might perform in more dynamic and realistic network environments.

\subsection{The Impact of Fast-HotStuff Countermeasure} \label{append:fhs-c}
In this section, we present an in-depth evaluation of the Fast-HotStuff countermeasure (FHS-C), focusing on its impact on throughput and latency compared to the original FHS protocol. The goal of these experiments is to quantify the trade-offs introduced by the additional dissemination step in FHS-C.

We conduct experiments under various network conditions to assess the performance of both FHS and FHS-C. The experiments are carried out with different network delay settings (0 ms, 10 ms, and 20 ms) and a varying number of nodes (from 4 to 64). Each experiment is repeated multiple times to ensure the reliability of the results.

\tabref{table:fhs-c-exp1} presents the throughput and latency of FHS and FHS-C under different network delays with 32 nodes. As expected, FHS-C introduces additional overhead due to the extra dissemination step. However, even with a network delay of 20 ms, FHS-C still achieves a throughput of 5880 transactions per second (decreases by 28\% compared to FHS) and a latency of 4007 ms (increases by 14\% compared to FHS), which are acceptable for many practical applications.

\begin{table}
    \caption{Throughput and latency of FHS and FHS-C protocols across various network delays.} 
    \label{table:fhs-c-exp1}
    \begin{subtable}{.25\textwidth}
      \centering
        \caption{Throughput (Tx/s)}
        \begin{tabular}{c|cc}
            \toprule[1pt]
            Delay & FHS & FHS-C  \\
            \midrule
            0ms & 30733 & 27515 \\
            10ms & 13904 & 10316 \\
            20ms & 8202 & 5880 \\ 
            \bottomrule[1pt]
        \end{tabular}
        \label{table:fhs-c-exp1-throughput}
    \end{subtable}%
    \begin{subtable}{.25\textwidth}
      \centering
        \caption{Latency (ms)}
        \begin{tabular}{c|cc}
            \toprule[1pt]
            Delay & FHS & FHS-C  \\
            \midrule
            0ms & 538 & 594 \\
            10ms & 1836 & 2193 \\
            20ms & 3503 & 4007 \\ 
            \bottomrule[1pt]
        \end{tabular}
        \label{table:fhs-c-exp1-latency}
    \end{subtable} 
\end{table}

\figref{fig:fhs-c-exp2-throughput} and \figref{fig:fhs-c-exp2-latency} illustrate the throughput and latency of FHS and FHS-C as the number of nodes increases. The experiment is conducted in a network environment with a delay of 10 ms. Both protocols exhibit a decrease in throughput and an increase in latency with more nodes. Nevertheless, FHS-C maintains a competitive performance level compared to FHS, indicating its effectiveness in preserving leadership democracy without significantly compromising overall system performance.

\begin{figure}[t]
    \centering
    \includegraphics[width=\linewidth]{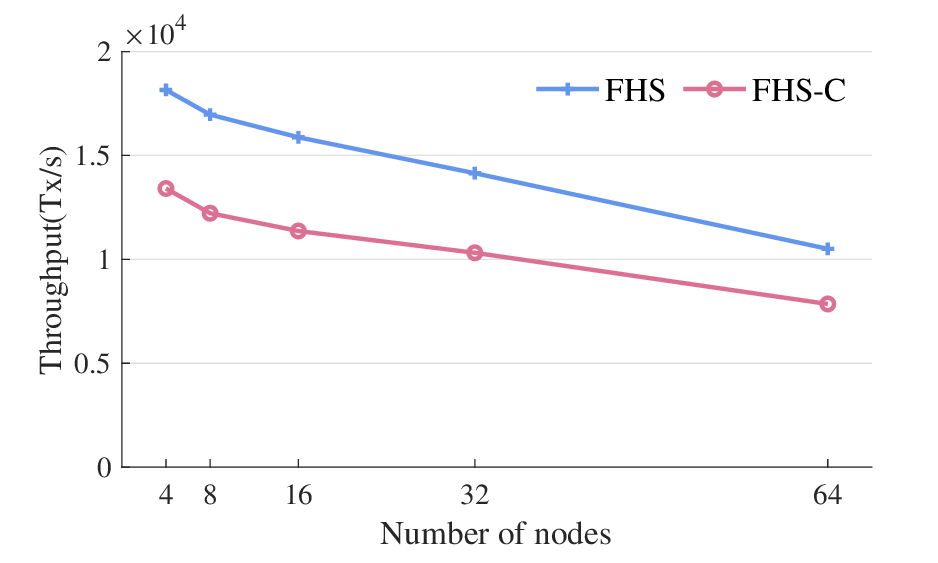}
    \caption{Throughput comparison between FHS and FHS-C as the number of nodes increases.}
    \label{fig:fhs-c-exp2-throughput}
\end{figure}

\begin{figure}[t]
    \centering
    \includegraphics[width=\linewidth]{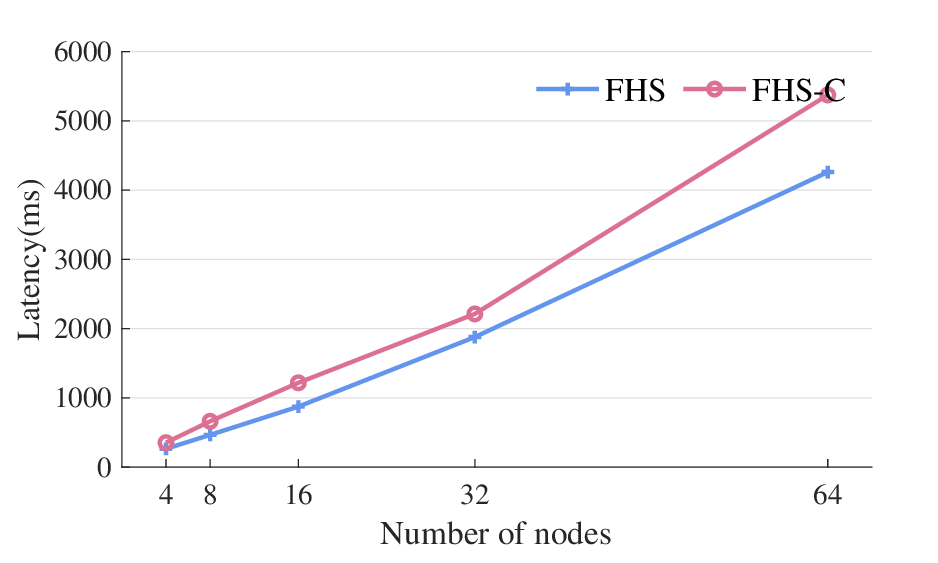}
    \caption{Latency comparison between FHS and FHS-C as the number of nodes increases.}
    \label{fig:fhs-c-exp2-latency}
\end{figure}

The experimental results demonstrate that while FHS-C introduces some overhead, it strikes a balance between leadership democracy and performance based on Fast-HotStuff.
These findings provide valuable guidance for researchers in understanding the practical implications of adopting FHS-C in different network environments.
}

\end{document}

%% file: macro.tex
\newcommand{\bheading}[1]{{\vspace{4pt}\noindent{\textbf{#1}}}}

\newcounter{note}[section]

\newcommand{\figref}[1]{\mbox{Fig.~\ref{#1}}}
\newcommand{\tabref}[1]{\mbox{Table~\ref{#1}}}
\newcommand{\appref}[1]{\mbox{Appendix~\ref{#1}}}
\newcommand{\ignore}[1]{}
\newcommand{\ssecref}[1]{\mbox{\S\ref{#1}}\xspace}

\newcommand{\lh}{\texttt{$l_h$}\xspace}
\newcommand{\la}{\texttt{$l_a$}\xspace}
\newcommand{\leader}{\texttt{$L$}\xspace}
\newcommand{\cnt}{$\texttt{$cnt$}$\xspace}

\newcommand{\metricOne}{chain quality\xspace}

\newcommand{\metricTwo}{cencorship\xspace}

\newcommand{\ie}{\textit{i.e.}\xspace}
\newcommand{\eg}{\textit{e.g.}\xspace}

\newcommand{\etal}{\textit{et al.}\xspace}

\newcounter{packednmbr}

\newenvironment{packeditemize}{
\begin{list}{$\bullet$}{
\setlength{\labelwidth}{0pt}
\setlength{\itemsep}{2pt}
\setlength{\leftmargin}{\labelwidth}
\addtolength{\leftmargin}{\labelsep}
\setlength{\parindent}{0pt}
\setlength{\listparindent}{\parindent}
\setlength{\parsep}{1pt}
\setlength{\topsep}{1pt}}}{\end{list}}

\newcounter{lessoncount}

\newtheorem{finding}{Finding}
\newtheorem{definition}{Definition}
\newtheorem{theorem}{Theorem}
\newtheorem{corollary}{Corollary}